\documentclass[letterpaper,showpacs,amsmath,amssymb,prb,aps,twocolumn,superscriptaddress]{revtex4}
\usepackage{graphicx}
\usepackage{color}
\def\rv{{\bf r}}
\def\pv{{\bf p}}
\def\Mv{{\bf M}}
\def\Hv{{\bf H}}

\def\Sv{{\bf \hat S}}
\def\Svdot{{\bf \dot{\hat S}}}
\def\eM{{\bf \hat e_M}}
\def\e3{{\bf \hat e_3}}
\def\ez{{\bf \hat e_z}}
\def\Dv{{\bf \delta \hat e_M}}
\def\tauv{{\boldmath $\hat \tau$}}

\def\up{\uparrow}
\def\down{\downarrow}
\def\lambdat{\stackrel{\leftrightarrow}{\lambda}}
\def\chit{\stackrel{\leftrightarrow}{\chi}}
\def\Sigmat{\stackrel{\leftrightarrow}{\Sigma}}
\def\ST{\stackrel{\leftrightarrow}{S}}
\def\epsilont{\stackrel{\leftrightarrow}{\epsilon}}
\def\onet{\stackrel{\leftrightarrow}{1}}
\def\Gammat{\stackrel{\leftrightarrow}{\Gamma}}

\begin{document}
\title {Gilbert damping and spin Coulomb drag in a magnetized electron liquid  with spin-orbit interaction}
\author {Ewelina M. Hankiewicz}
\affiliation{Department of Physics and Astronomy, University of Missouri, Columbia, Missouri 65211, USA}
\author {Giovanni Vignale}
\affiliation{Department of Physics and Astronomy, University of Missouri, Columbia, Missouri 65211, USA}
\author {Yaroslav Tserkovnyak}
\affiliation{Department of Physics and Astronomy, University of California, Los Angeles, California 90095, USA}
\date{\today}
\begin{abstract}
We present a microscopic calculation of the Gilbert damping constant for the magnetization of
a two-dimensional spin-polarized electron liquid in the presence of intrinsic spin-orbit interaction.
 First we show that the Gilbert constant can be expressed in terms of the auto-correlation function of the spin-orbit
  induced torque.   Then we specialize to the case of  the Rashba spin-orbit interaction and we show that the Gilbert
   constant in this model is related to the spin-channel conductivity.  This allows us to study the  Gilbert damping
    constant in different physical regimes, characterized by different orderings of the relevant energy scales -- spin-orbit
     coupling, Zeeman coupling, momentum relaxation rate, spin-momentum relaxation rate, spin precession frequency -- and
     to discuss its behavior in various limits.  Particular attention is paid to electron-electron interaction effects,
     which enter the spin conductivity and hence the Gilbert damping constant via the spin Coulomb drag coefficient.
\end{abstract}
\pacs{}
\maketitle

\section{Introduction}\label{Introduction}

The Gilbert constant characterizing the damping of magnetization
precession is one of the important phenomenological parameters
that describe the collective magnetization dynamics of
ferromagnets.\cite{Gilbert55,Gilbert04,landauBOOKv9}. It is an
essential input of  the  Landau-Lifshitz-Gilbert equation (LLG)
for magnetization dynamics\cite{landauBOOKv9} and as such is
widely used in the analysis of magnetization reversal processes,
which are crucial to magnetic recording technologies
\cite{Yaroslav05}. Recently, hybrid systems of ferromagnets and
normal metals have also attracted considerable attention, in the
context of the enhancement of Gilbert damping at the
ferromagnet-normal metal
interfaces.\cite{tserkovPRL02sp,tserkovPRB02sp,Yaroslav05}
 Despite tremendous efforts to elucidate the nature of Gilbert damping in bulk ferromagnets,
 however, the microscopic processes responsible for the observed ferromagnetic relaxation in real materials
 are still not fully understood.
 The matter is complicated by the potential relevance of a number of different mechanisms
 involving eddy currents, magnetoelastic coupling, two-magnon and nonlinear multimagnon processes,
 extrinsic and intrinsic spin-orbit (SO) coupling of itinerant electrons.\cite{heinrichPSS67,Korenman72,lutovinovJETP79,solontsovPLA93,suhlIEEEM98,heinrichIEEEM02,kunesPRB02,dobinPRL03,sinovaPRB04,Tserkovnyak04,koopmansPRL05,steiaufPRB05,tserkovPRB06,kohnoCM06}
Without delving into a detailed discussion of various points of
view, we note that a certain level of consensus has been reached
in the recent theoretical literature about the central importance of
SO interaction of some form in conducting ferromagnets, although
with a limited and rather indirect experimental support at
present.\cite{mizukamiPRB02,ingvarssonPRB02}

Even restricting the attention to the SO-based mechanisms of Gilbert damping,
however, the myriad of relevant energy scales (namely, ferromagnetic resonance frequency,
 ferromagnetic exchange energy, intrinsic SO splitting, impurity scattering rate, and spin
 dephasing due to magnetic or SO disorder) has led different authors to make qualitatively
  different predictions for the Gilbert damping, for example, with regard to its dependence on
   the disorder strength.\cite{heinrichPSS67,kunesPRB02,sinovaPRB04,Balents04,Tserkovnyak04,kohnoCM06}.
    Furthermore,  electron-electron ($e-e$) interactions have mainly been discussed in the mean-field spirit,
    as the source of the exchange field in itinerant-electron ferromagnets.

In this paper, we set off to formulate a microscopic theory of the Gilbert damping constant,
which we then apply to a simple model, where the competition of different energy scales can be studied,
 comparing various points of view and also generalizing and extending the results existing in the literature.
 We wish to consider the interplay of magnetic spin splitting, intrinsic SO strength, disorder scattering, as well as the strength of $e-e$ interactions in a single self-contained model without introducing any phenomenological dephasing or relaxation parameters. In particular, we find an intricate relation between the spin-drag correlations induced by $e-e$ interactions and the Gilbert damping in two-dimensional electron liquids. We hope our discussion will bring the community a step closer to understanding the relevant microscopic mechanisms responsible for the Gilbert damping observed in conducting ferromagnets and spin-polarized systems.

SO interactions, especially in narrow-band semiconductors,
 have recently received a great deal of attention in the context of spin-based electronics
 (i.e., spintronics). In particular, theoretical proposals to manipulate spins by means of
  intrinsic SO interactions, without the use of magnetic fields or magnetic materials,
  especially in the so-called spin Hall configuration,\cite{Murakami03,Sinova04} have unleashed
  a wave of theoretical research as well as experimental efforts to measure the effect.\cite{Kato04,Wunderlich05,sihPRL06}
  It is worthwhile noting that these activities have shown the need to better understand the
  fundamental aspects of intrinsic SO coupling and its experimental manifestations. The role of $e-e$ interactions,
  furthermore, remains a relatively unexplored territory with regard to its interplay with interesting topological
  properties brought about by the SO coupling, which lie beyond the conventional
   theories of electron liquids. In addition, the recent discovery of the Aharonov-Casher phase\cite{Konig06}
   in narrow-band semiconductors shows that a study of the combined response to magnetic and SO fields
   in a regime of weak and strong SO interactions is of urgent importance.
   It is closely related to the topic considered here, i.e., the spin response and relaxation
   in magnetized system with SO interactions.

In this paper, we study an interacting electron liquid magnetized
by an external magnetic field in the presence of SO interactions.
First, by comparing the macroscopic Landau-Lifshitz-Gilbert (LLG)
theory with a microscopically derived expression for the
transverse spin susceptibility, we find the following relationship
between the Gilbert damping constant and the torque-torque correlator (see
Eq.~32 below):
\begin{equation}\label{BasicRelation}
\lambda=-\frac{g}{M_0V}\lim_{\omega\to0}\Im
m\frac{\langle\langle\hat\tau;\hat\tau\rangle\rangle_{\omega}}{\omega}
\end{equation}
where $g$ is the gyromagnetic ratio, $\lambda$ is the
dimensionless Gilbert coefficient, $M_0$ is the equilibrium
magnetization, $V$ the volume of the system, and $\hat{\tau}$ the
torque induced by SO interactions.  To derive
Eq.~(\ref{BasicRelation})  we must neglect certain contributions
of  order  higher than the fourth in the strength of the
spin-orbit coupling: we will argue  that this procedure is
justified, not only at zero frequency but also at frequencies
close to the ferromagnetic resonance.

Next we focus on the specific case of a two-dimensional electron gas (2DEG) with Rashba SO interactions induced by a uniform electric field perpendicular to the plane of the electron gas.   In this case, we show that the Gilbert damping constant can be expressed in terms of the spin-channel conductivity:  the same result holds in a three-dimensional electron gas, provided the electric field responsible for the spin-orbit interaction is uniform in space.    For the isotropic case, i.e., when the magnetic field, the magnetization, and the SO-inducing electric field are all perpendicular to the direction of the 2DEG, the result is
\begin{equation}
\lambda=\frac{\bar\alpha^2E_F\tau}{\hbar p}\frac{\Re e\sigma_{s\parallel}(0)}{\sigma_D}\,,
\label{scc}
\end{equation}
where $E_F$ is the Fermi energy, $\tau$ is the momentum  relaxation time due to electron-impurity scattering, $\bar\alpha$ is the Rashba SO-coupling strength normalized by the Fermi velocity $v_F$, $p$ is the degree of spin polarization, $\sigma_D$ is the Drude conductivity, and $\sigma_{s\parallel}(0)$ is the longitudinal spin-channel conductivity.

The problem now shifts to the calculation of the spin-channel conductivity. We show that this can be done exactly (subject to the usual weak-disorder assumption $E_F\tau\gg 1$) in the isotropic non-interacting case, with ladder vertex corrections playing an essential role in ensuring the correct behavior in the limit of strong SO coupling.  $e-e$ interactions  can then be introduced using the formalism described in Refs.~\onlinecite{Amico00,Amico02}. The final result for the isotropic case at zero frequency has the form:
\begin{equation}\label{lambda_general}
\lambda=\frac{p\bar \alpha^2
E_F\tau}{\hbar(p^2+\bar\alpha^2)}\frac{1+\frac{1}{2}(\Omega\tau)^2(1+\cos^2\delta)}{\cos^2\delta+\frac{1}{4}(\Omega\tau)^2(1+\cos^2\delta)^2}\frac{1+p^2\gamma\tau}{1+\gamma\tau}\,,
\end{equation}
where $\Omega$ is the spin splitting of electronic states due to the combined action of the external magnetic field and the SO effective magnetic field,
$\cos\delta$ is the ratio of the external magnetic field to the total effective field $\Omega$, and $\gamma$ is the spin-drag coefficient.
This expresses the Gilbert damping as a function of disorder strength, magnetic field, SO and $e-e$ interactions.

One of the gross features of Eq.~(\ref{lambda_general}) is a rough scaling of $\lambda$ with $\tau$, which is particularly evident
in the clean limit $\tau\to\infty$. We will see, however, that Eq.~(\ref{lambda_general}) describes  in general
 a nonmonotonic dependence of $\lambda$ on the scattering rate, due to the interplay between $1/\tau$ and other energy scales. For example, in the regime of strong $e-e$ interactions and weak polarizations, $1<\gamma\tau<1/p^2$, the Gilbert damping $\lambda$ becomes independent of $\tau$ and scales as $1/\gamma$.

It is important to note that Eq.~(\ref{lambda_general}) holds in the low-frequency limit, $\omega\to0$,
which means in practice $\omega$ much smaller than all the other energy scales. For this reason the divergence of
 $\lambda$ as $1/\tau\to0$ in Eq.~(\ref{lambda_general}), which was obtained after first taking the $\omega\to0$ limit,
 is not physically consequential: taking the $1/\tau\to0$ limit with a fixed $\omega$ results in a vanishing damping as
 it should.\cite{Tserkovnyak04}.
 This is clearly seen by extending the calculation of $\lambda$ to finite frequency, which can be done with
 little extra effort in the $\omega \ll E_F$ regime.  This important extension of Eq.~(\ref{lambda_general}) is presented in
 full in Eq.~59 below. Among other modifications we get a factor $\frac{1}{1+(\omega \tau)^2}$
  to multiply Eq.~(\ref{lambda_general}),  ensuring the correct behavior of
$\lambda$  for $\tau \to \infty$ at finite $\omega$.

The rest of the paper is organized as follows. In Sec.~\ref{LLGequation}, we
discuss the LLG phenomenology, which is compared to the microscopic calculation in Sec.~\ref{MicroscopicTheory},
allowing us to express the Gilbert damping constant in terms of the torque-torque correlator.
In Sec.~\ref{IsotropicCase},
 we present the calculation of the Gilbert damping constant in the presence of the Rashba SO
  and $e-e$ interactions for
 an isotropic case, i.e., when the magnetic field is perpendicular to the 2DEG plane.
  In Sec.~\ref{AnisotropicCase}, the same model is treated for the anisotropic case,
  where the induced magnetization is tilted away from the direction perpendicular to the 2DEG.
  Our summary, conclusions and speculations are presented in Sec.~\ref{Summary}.  Technical details of the calculations are supplied in the Appendixes.

\section{LLG equation}\label{LLGequation}

Studies of the magnetization dynamics in ferromagnetic materials often start from
a phenomenological equation of motion for the magnetization
\begin{equation}
\Mv(\rv,t)=-g\langle\mathbf{\hat{s}}(\rv,t)\rangle\,,
\end{equation}
where $\langle\mathbf{\hat{s}}(\rv,t)\rangle$ is the expectation value of the spin-density operator at position $\rv$ and $g$ is the gyromagnetic ratio: for free electrons in vacuum, $g=e/mc$  where $e$ is the absolute value of the electron charge. An important assumption in the LLG phenomenology is that the {\it magnitude} of the magnetization is fixed at a constant value $M_0$ (determined by the minimization of the free energy), i.e., we have
\begin{equation}
\Mv = M_0 \eM\,,
\end{equation}
where $\eM$ is the unit vector specifying the {\it direction} of the magnetization.
  The physical reasoning\cite{landauBOOKv9} underlying this picture assumes a very large (essentially infinite) longitudinal spin stiffness which prevents changes in the magnitude of $\Mv$, while one focuses on the softer transverse modes  -- essentially ferromagnetic Goldstone modes with a finite gap in the presence of intrinsic or extrinsic anisotropies. The direction of the magnetization $\eM$ thus provides the relevant degrees of freedom, whose time evolution is determined by the LLG equation
\begin{equation}\label{LLG}
\frac{\partial \eM}{\partial t}=g \Hv_{\rm eff}\times \eM +  \lambdat\cdot\,\eM\times\frac{\partial \eM}{\partial t}\,,
\end{equation}
where $\Hv_{\rm eff}$ is the effective magnetic field defined as the functional derivative of the equilibrium free-energy {\it density} $f(\Mv)$, regarded as a functional of the magnetization $\Mv$, with respect to its argument:
\begin{equation}\label{Heff}
\Hv_{\rm eff}(\rv,t)=-\frac{\delta f(\Mv)}{\delta \Mv(\rv,t)}\,.
\end{equation}
$\lambdat$ is a symmetric $2 \times 2$ matrix acting in the two-dimensional space
perpendicular to the magnetization, and the dot denotes the usual matrix product [note that the possible antisymmetric component of $\lambdat$ would only lead to a renormalization of the gyromagnetic ratio  in Eq.~(\ref{Heff}).]
 The first term on the right-hand side (r.h.s.) of this equation produces a coherent
 precessional motion of the magnetization, conserving the total free energy,
  while the second term, known as the Gilbert damping, is responsible for the relaxation toward equilibrium, i.e., the lowest free-energy state.
  Both terms vanish at equilibrium.
  Notice that the LLG equation is not restricted to the linear-response regime, as $\eM$ is allowed to wander
   away arbitrarily from the equilibrium orientation.

The primary focus of this paper is on the microscopic calculation of the Gilbert damping matrix $\lambdat$.
   We concentrate exclusively on homogeneous systems, that is the situation in which $\Mv$ is independent
   of  position and depends only on time.  In such a case, it is immediately obvious that any damping must
   arise from the coupling between the spin and the orbital degrees of freedom.  This is so since
   the nonrelativistic $e-e$ interaction is invariant under a uniform rotation of all
   the spins, and thus cannot be involved in the damping of such a rotation.
   It will therefore be essential to include the SO coupling in our microscopic calculation,
   and we will do this starting from the simplest model, and then try to generalize our conclusions to
    more realistic situations. Note that although in our model we consider SO coupling stemming from
    the relativistic corrections to the Schr{\"{o}}dinger equation, a kind of SO coupling would
    also be possible in the presence of an inhomogeneous crystal magnetic field.

\begin{figure}[thb]
\vskip 0.27 in
\includegraphics[width=3.3in]{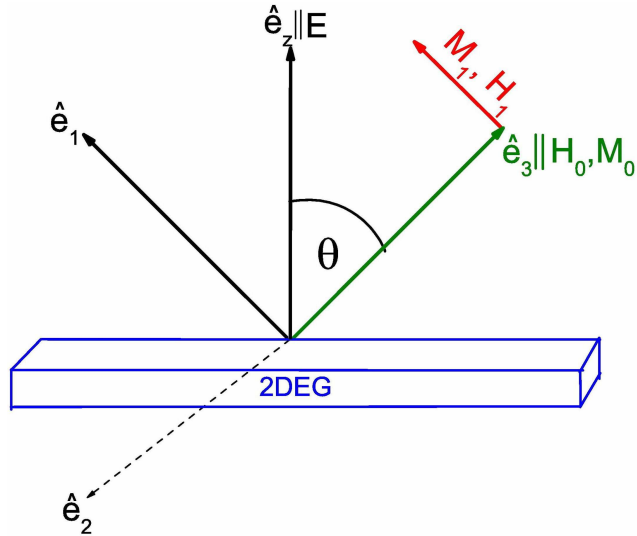}
\caption{The directions of the magnetic and Rashba fields in
relation to the 2DEG plane, the model which we discuss in  Section
IV. Also, visualization of directions of equilibrium and induced
magnetization presented in Eqs.~8-14.} \label{Graph1}
\end{figure}

We can calculate $\lambdat$ by applying Eq.~(\ref{LLG}) to a
linear-response problem and comparing the resulting
susceptibility with the susceptibility obtained from the
Kubo formula.  Our linear-response problem is depicted in Fig.~\ref{Graph1}.
We perturb the system with a small magnetic field $\Hv_1$
perpendicular to the direction of the {\it equilibrium}
magnetization, which we denote by $\e3$ to distinguish it from the
actual direction of the magnetization $\eM$.  $\Hv_1$ is uniform
in space and periodic in time, with an angular frequency $\omega$
(this field must not be confused with the static field $\Hv_0$
which may be present at equilibrium).   In response to $\Hv_1$, the
direction of the magnetization becomes time dependent and will be
written as
\begin{equation}\label{M}
\eM (t) = \e3+ \Dv(t)\,,
\end{equation}
where $\Dv$ is a small vector perpendicular to $\e3$, which
therefore does not change the normalization of $\eM$ to first
order. Adding the external field $\Hv_1$ to $\Hv_{\rm eff}$ and
substituting Eq.~(\ref{M}) into Eq.~(\ref{LLG}), we obtain the equation
of motion for  $\Dv$ to first order in $\Hv_1$:
\begin{equation}\label{linearized_LLG}
\frac{\partial  \Dv}{\partial t} =g(\Hv_1+\Hv_{\rm eff,1})\times
 \e3 +  \lambdat \cdot \,\e3\times\frac{\partial \Dv}{\partial t}\,.
\end{equation}
To the same order, the effective field is
\begin{eqnarray}\label{Heff1}
\Hv_{\rm eff,1}&=&-\frac{1}{M_0^2}\left.\frac{\partial^2f(\Mv)}{\partial \eM \partial
\eM}\right \vert_{\e3} \cdot  M_0 \Dv\nonumber\\
&=& - \ST \cdot M_0 \Dv\,,
\end{eqnarray}
where we have defined the {\it transverse} spin stiffness matrix
\begin{equation}\label{stiffness}
\ST \equiv \frac{1}{M_0^2} \left.\frac{\partial^2f(\Mv)}{\partial \eM \partial
\eM}\right \vert_{\e3}\,,
\end{equation}
a $2\times 2$ matrix in the plane perpendicular to $\e3$.

Upon taking a Fourier transform with respect to time, the linearized LLG equation can be rewritten as follows:
\begin{equation}\label{linearized_LLG2}
\left[-i\omega(\onet+\epsilont \cdot \lambdat)+g M_0
\epsilont\cdot\ST\right]\cdot \Dv = g \epsilont\cdot\, \Hv_1\,,
\end{equation}
where $\onet$ is the $2 \times 2$ identity matrix and
\begin{equation}\label{Levi-Civita}
\epsilont\equiv\left( \begin{array}{cc}
  0 & 1\\
  -1 & 0
\end{array}\right)\,,
\end{equation}
$\epsilont^{-1}=-\epsilont$.
The transverse spin susceptibility $\chit$ connects the transverse magnetization
$\Mv_1 = M_0\Dv$ to the external field $\Hv_1$:
\begin{equation}\label{defchi}
\Mv_1(\omega) = \chit (\omega) \cdot \Hv_1(\omega)\,.
\end{equation}
Comparing this with Eq.~(\ref{linearized_LLG2}), we easily obtain
\begin{equation}\label{susceptibility}
\chit^{-1}(\omega) = \frac{i\omega(\epsilont-\lambdat)}{gM_0}+\ST\,,
\end{equation}
which is the most general form of the anisotropic transverse response, to the first order in frequency. Notice that $\chit^{-1}(0)=\ST$, showing that the transverse stiffness is indeed the inverse of the static spin
susceptibility.
In an isotropic system, $\ST$ vanishes (the free energy does not depend on the direction of the magnetization).
If a static external field  $\Hv_0 = H_0 \e3$
is applied to such a system, then the stiffness matrix becomes $\ST=(H_0/M_0)\onet$.
Anisotropies due to SO interactions will show up as additional contributions to $\ST$.

Needless to say, the theory described in this section is valid
only at low frequencies (in comparison to relevant microscopic energy scales), when the frequency expansion of the inverse susceptibility, such as Eq.~(\ref{susceptibility}), can be truncated at the linear term.

\section{Microscopic theory}\label{MicroscopicTheory}

A microscopic expression for the homogeneous transverse spin susceptibility is given by the Kubo formula
\begin{equation} \label{KuboFormula}
\chi_{ij}(\omega) = \frac{g^2}{V} \left[\frac{i}{\hbar}
\int_0^\infty dt \langle [ \hat S_i(t), \hat S_j(0)] \rangle e^{i
(\omega+i0^+) t}\right]\,,
\end {equation}
where $\hat S_i(t)$ is the Heisenberg operator of the total spin
of the system, the angular bracket denotes the equilibrium
average, $V$ is the volume of the system, and $i$, $j$ are
Cartesian indices in the plane perpendicular to the magnetization
(they can take up the values $1$ or $2$). The lengthy expression
inside the square brackets in the above equation (i.e., the
retarded spin response function) will be abbreviated  from now on
as $-\langle \langle \hat S_i;\hat S_j \rangle\rangle_\omega$, so
we have
\begin{equation} \label{KuboFormula2}
\chi_{ij}(\omega) = -\frac{g^2}{V} \langle \langle \hat S_i;\hat
S_j \rangle\rangle_\omega\,.
\end {equation}
(The magnetization is related to the average spin by
$\Mv=-g\langle {\bf \hat S} \rangle/V$.)
 Notice that this expression is dimensionless in three dimensions and has the units of length in two spatial dimensions (in the cgs units).

In view of the fact that Gilbert damping arises ultimately from
torques on the individual spins, it is convenient to rewrite
Eq.~(\ref{KuboFormula2})
 in terms of the time derivatives of $\Sv$.
 The time derivative of the spin is conveniently written as the sum of two parts:  a free precession
term at frequency $\omega_0$ about the magnetization axis $\e3$ and a residual torque  \tauv\ due
to the SO interactions:
\begin{equation}\label{spintorque}
\Svdot = \omega_0 \e3 \times \Sv + \mbox{\tauv}\,.
\end{equation}
Here, $\omega_0=gH_0'$, where $H_0'$ is the projection of the static magnetic field along $\e3$, while \tauv\ is a model-dependent torque, which is proportional to the spin-orbit coupling constant.

 It is very convenient at this point
to rewrite Eq.~(\ref{KuboFormula2}) in terms of the
 $\langle \langle \mbox{\tauv};\mbox{\tauv}\rangle\rangle$ correlation function.
  The fact that $\mbox{\tauv}$  explicitly contains the spin-orbit coupling constant
  $\alpha$ will allow us  to calculate the correlator,  to order $\alpha^2$,
   without including the spin-orbit interaction in the hamiltonian: this will be a major simplification.  Furthermore, this transformation will enable us to connect the Gilbert constant  to the spin conductivity,  which is particularly helpful for the inclusion of $e-e$ interactions.

 To express the spin susceptibility in terms of the
 $\langle \langle \mbox{\tauv};\mbox{\tauv}\rangle\rangle$ correlator we make use of the identity
\begin{eqnarray}\label{eom}
\langle \langle \hat A;\hat B \rangle \rangle_\omega
&=&-\frac{\langle \langle \dot  {\hat A};\hat B \rangle
\rangle_\omega}{i \omega} +\frac{\langle [\hat A,\hat B]
\rangle}{\hbar \omega}\nonumber\\
&=&+\frac{\langle \langle {\hat A};\dot{\hat B} \rangle
\rangle_\omega}{i \omega} +\frac{\langle [\hat A,\hat B]
\rangle}{\hbar \omega}\,,
\end{eqnarray}
where
\begin{equation}\label{eq_of_motion}
\dot{\hat A}=\frac{1}{i\hbar}[\hat A,\hat H]
\end{equation}
is the time derivative of $\hat A$  and $\hat H$ is the unperturbed Hamiltonian.

Combining Eqs.~(\ref{eom}) and~(\ref{spintorque}), we easily obtain
\begin{eqnarray}\label{chi1}
&&-i\omega\langle\langle \hat S_ i;\hat
S_j\rangle\rangle_\omega=\langle\langle \dot{\hat S}_i;\hat
S_j\rangle\rangle_\omega -\frac{i}{\hbar}\langle[\hat S_i,\hat
S_j]\rangle\nonumber\\ &=&\epsilon_{ij}\langle \hat
S_3\rangle-\omega_0\epsilon_{ik}\langle\langle \hat S_k;\hat
S_j\rangle\rangle_\omega +\langle\langle \hat \tau_i;\hat
S_j\rangle\rangle_\omega\,,
\end{eqnarray}
where a sum over the repeated index ($k$ in this case) is
understood. Solving for $\langle\langle \hat S_ i;\hat
S_j\rangle\rangle_\omega$, we get
\begin{equation}\label{chi4}
\langle\langle \hat S_ i;\hat
S_j\rangle\rangle_\omega=-\frac{i\omega
\delta_{ik}+\omega_0\epsilon_{ik}}{\omega^2_0-\omega^2}\left[\epsilon_{kj}\langle
\hat S_3\rangle + \langle\langle \hat \tau_k;\hat
S_j\rangle\rangle_\omega\right]\,.
\end{equation}
In the absence of the SO coupling, the torque \tauv\ vanishes and the second term on the r.h.s.
of Eq.~(\ref{chi4}) is absent.  Then the susceptibility is easily found to have the form
\begin{equation} \label{chi0}
\chi^{(0)} _{ij}(\omega) = g M_0 \frac{-i\omega \epsilon_{ij}+\omega_0\delta_{ij}}{\omega_0^2-\omega^2}~.
\end{equation}
Notice that its inverse
\begin{equation}\label{chi0_inverse}
[\chit^{(0)}]^{-1}(\omega)=\frac{i\omega \epsilont+\omega_0\onet}{g M_0}
\end{equation}
is in a perfect agreement with Eq.~(\ref{susceptibility}), if in the latter we  set $\lambdat=0$ and $\ST = (\omega_0/gM_0) \onet$.

Upon applying again Eq.~(\ref{eom}) to $\langle\langle \hat \tau_k;\hat S_j\rangle\rangle_\omega$, we get
\begin{equation}\label{tau_S}
\langle\langle \hat \tau_k; \hat S_j\rangle\rangle_\omega =
\left[\frac{i}{\hbar}\langle[\hat \tau_k, \hat S_l]\rangle+\langle\langle \hat  \tau_k; \hat \tau_l\rangle\rangle_\omega\right]\frac{i\omega \delta_{lj}+\omega_0\epsilon_{lj}}{\omega^2_0-\omega^2}\,,
\end{equation}
Putting this in Eq.~(\ref{chi4}) and making use of Eqs.~(\ref{KuboFormula2}) and~(\ref{chi0}),
 we arrive after some algebraic manipulations at the following compact formula for the spin susceptibility matrix:
\begin{equation}\label{chi_matrix}
\chit(\omega)=\chit^{(0)}(\omega) + \chit^{(0)}(\omega)\cdot\frac{\epsilont\cdot \Gammat (\omega)\cdot\epsilont}{M_0^2 V}\cdot \chit^{(0)}(\omega)\,,
\end{equation}
where  the matrix $\Gammat(\omega)$ is defined as follows:
\begin{equation}\label{Gamma}
\Gamma_{ij}(\omega)=\langle\langle \hat \tau_i;\hat
\tau_j\rangle\rangle_\omega + \frac{i}{\hbar}\langle[\hat
\tau_i,\hat S_j]\rangle\,.
\end{equation}
The main dynamical quantity on the r.h.s. here is the torque-torque response
 function $\langle\langle \hat \tau_i;\hat \tau_j\rangle\rangle_\omega$.
 The rest is ground-state averages that we will not be interested in.

We can now express the Gilbert damping and stiffness matrices, $\lambdat$ and $\ST$,
which appear in the phenomenological equation~(\ref{susceptibility}), in terms of microscopic response functions. First of all, consider the formal limit of weak SO interactions.
In this limit, $\Gammat$, which is proportional to the square of the SO coupling constant,
is assumed to give a small correction to the susceptibility (we will comment below on
the conditions of applicability of this regime).  Upon inverting  Eq.~(\ref{chi_matrix}) to first order in $\Gammat$,
we obtain
\begin{equation}\label{chi_matrix.inverse}
\chit^{-1}(\omega) \simeq  [\chit^{(0)}]^{-1}(\omega) -\frac{\epsilont\cdot \Gammat(\omega)\cdot\epsilont}{M_0^2 V}\,.
 \end{equation}
Comparing this with Eq.~(\ref{susceptibility}), and taking into account Eq.~(\ref{chi0_inverse}), we get
\begin{equation}\label{S.spinorbit}
\ST_{so}  = -\frac{\epsilont\cdot \Gammat(0)\cdot\epsilont}{M_0^2V}
\end{equation}
and
\begin{equation}\label{GilbertMatrix}
\lambdat = \frac{g}{M_0V} \lim_{\omega \to 0}
\left.\Im m \frac{\epsilont\cdot \Gammat (\omega)\cdot\epsilont}{\omega}\right\vert_{sym}\,,
\end{equation}
where the subscript $sym$ denotes the symmetric part of a matrix.  In the first of these two equations, $\hat S_{so}$ is the SO contribution to the stiffness matrix, i.e.,  $\ST_{so} \equiv \ST - (H_0'/M_0) \onet$, which reflects SO induced magnetic anisotropy.  The second equation is the main result at this point.  It expresses the Gilbert damping matrix in terms of the zero-frequency slope of the spectrum of the torque-torque response function. Notice that this spectrum has no contribution from the second term on the r.h.s. of Eq.~(\ref{Gamma}), which  is purely real (the commutator of two hermitian operators is anti-hermitian).  In general,  the imaginary part of $\epsilont\cdot \Gammat (\omega)\cdot\epsilont$  will have both symmetric and antisymmetric components: the latter has been interpreted as a Berry curvature correction\cite{qianPRL02} to the adiabatic spin dynamics [also, see the note below Eq.~(\ref{Heff}), where we point out that inclusion of the antisymmetric component would lead to a renormalization of the $g$-factor].  The symmetric component is purely diagonal in the limit of weak spin-orbit coupling because in this limit the system is isotropic under rotations about the $3$ axis.  (In general, one can still diagonalize the symmetric component of the spectrum by a suitable choice of the axes in the $1-2$ plane -- a transformation that does not affect the antisymmetric component. We will not consider this complication here).    Thus we conclude that the Gilbert damping matrix is a diagonal matrix of the form
\begin{equation}\label{Lambda.Matrix}
\lambdat\equiv\left( \begin{array}{cc}
\lambda_1 & 0\\
  0 & \lambda_2
\end{array}\right)\,,
\end{equation}
where
\begin{equation}\label{Gilbert11}
\lambda_1 =-\frac{g}{M_0V} \lim_{\omega \to 0}  \Im m \frac{\langle\langle
\hat \tau_2; \hat \tau_2\rangle\rangle_{\omega}}{\omega}
\end{equation}
and
\begin{equation}\label{Gilbert22}
\lambda_2 =-\frac{g}{M_0V} \lim_{\omega \to 0}  \Im m \frac{\langle\langle
\hat \tau_1; \hat \tau_1\rangle\rangle_{\omega}}{\omega}\,.
\end{equation}

The above  formulas  have been derived from a first order
expansion in $\Gammat$, which is justified for $\omega \ll \omega_0$ when the spin-orbit interaction is weak.  At first sight, however, the approximation seems to break down completely when $\omega \sim \omega_0$  (the interesting region of the ferromagnetic resonance)  because $\chit^{(0)}$ has a pole at $\omega = \omega_0$.    But this conclusion is too hasty.
To show this, we first notice that the {\it exact} inversion of $\chit$ from Eq.~(\ref{chi_matrix}) gives

\begin{equation}\label{exact.inverse}
\chit^{-1}(\omega) =  [\chit^{(0)}]^{-1}(\omega) -\Sigmat(\omega)\,
 \end{equation}
 where the ``self-energy" $\Sigmat(\omega)$ is given by
 \begin{equation}\label{selfenergy}
 \Sigmat (\omega) =  \frac{\epsilont\cdot \Gammat(\omega)\cdot\epsilont}{M_0^2 V}\left[\onet+ \chit^{(0)}(\omega)\cdot\frac{\epsilont\cdot \Gammat (\omega)\cdot\epsilont}{M_0^2 V}\right]^{-1}\,.\\
 \end{equation}

For $\omega \ll \omega_0$,  $\chit^{(0)}(\omega)$ reduces to $S_0^{-1}=\frac{gM_0}{\omega_0}$, which is just the inverse of the stiffness to zero-th order in the spin-orbit interaction.  Then, from the above formulas and making use of Eq.~(\ref{S.spinorbit})  we see that the zero-frequency Gilbert matrix (proportional to $\Im m \Sigmat (\omega)/\omega$)  is given by Eq.~(\ref{GilbertMatrix}) times a renormalization  factor
\begin{equation}\label{renormalization}
\left(\onet - \frac{\ST_{so}}{S_0} \right)^{-1}~.
\end{equation}
We will estimate this renormalization below for a specific model, and find it to be very small.

An important observation is that, unlike the function $\chit^{(0)}(\omega)$, the self-energy is well-behaved at $\omega=\omega_0$.   Indeed from Eq.~(\ref{chi_matrix}) we see that  $\Gammat(\omega)$ must vanish at $\omega_0$ in order to cancel the pole of $\chit^{(0)}$ and replace it with a roughly Lorentzian peak at the true ferromagnetic resonance.    Furthermore,  the rate at which $\Gammat(\omega)$ vanishes for $\omega$ tending to $\omega_0$ must be such that the denominator of Eq.~(\ref{selfenergy}), i.e.,   $\left[\onet+ \chit^{(0)}(\omega)\cdot\frac{\epsilont\cdot \Gammat (\omega)\cdot\epsilont}{M_0^2 V}\right]$, must also vanish at $\omega_0$ -- otherwise the pole at $\omega_0$ would survive.  (Note: we say that an operator vanishes or has a pole in the sense that its projection along the eigenvector of the ferromagnetic resonance, $(1,i)$ or $(1,-i)$, has a zero or a pole).   Therefore $\Sigmat(\omega_0)$ is finite and amenable to a perturbative treatment in which we retain only the leading term $\Sigmat (\omega) \simeq \frac{\epsilont\cdot \Gammat(\omega)\cdot\epsilont}{M_0^2 V}$.   This leads to the more general formula
\begin{equation}\label{GilbertMatrixOmega}
\lambdat (\omega) = \frac{g}{M_0V}
\left.\Im m \frac{\epsilont\cdot \Gammat (\omega)\cdot\epsilont}{\omega}\right\vert_{sym}\,,
\end{equation}
which is applicable at finite frequency, as well as zero frequency.  Furthermore, we will show below that this approximation for $\Sigmat$ generates results that are in perfect agreement, in the
non-interacting case, with the direct diagrammatic calculation of the spin susceptibility, even when $\alpha$ is {\it not} small.

\section{Spin response of a Rashba 2DEG} \label{ModelCalculation}

In this section we introduce a simple model that allows us to
calculate the torque-torque response function and hence the
Gilbert damping and the static stiffness from first principles.
The model is illustrated in Fig.~1 and consists of a
two-dimensional electron gas (2DEG) with a linear SO
interaction induced by an electric field along the $z$ axis
perpendicular to the plane of the electrons, and a magnetic field
$\Hv=H_0\e3$ where the unit vector $\e3$ forms an angle $\Theta$
with the $z$ axis and lies in the $z-x$ plane.   The hamiltonian
for the model is
\begin{eqnarray}\label{modelHamiltonian}
\hat H &=& \sum_n \left[\frac{\hat p_n^2}{2m}+V(\hat \rv_n)+\omega_0 \Sv_n \cdot \e3
-\frac{\alpha}{\hbar} (\ez \times \pv) \cdot \Sv_n  \right]\nonumber\\&+&\frac{1}{2}
 \sum_{n \neq n'}\frac{e^2}{\epsilon_b \vert \rv_n - \rv_{n'}\vert}~,
\end{eqnarray}
where $\hat \pv_n$, $\hat \rv_n$, and $\Sv_n$ are, respectively,
the momentum, the position, and the spin operators of the $n$-th
electron, $V(\rv)$ is the random electron-impurity potential,
$\alpha$ is the Rashba velocity, which controls the strength of
the SO coupling  ($\alpha$ is proportional to the electric
field in the $z$ direction), $m$ is the effective mass of the
electrons and $\epsilon_b$ is the background dielectric constant.
The explicit form for $\alpha$ in a typical direct gap
semiconductor (say GaAs) is
\begin{equation}\label{defalpha}
\alpha = v_F (eEa_B)\left(\frac{2P^2}{3m}\right) \left(\frac{1}{E_g^2}-\frac{1}{(E_g+\Delta_{so})^2}
\right) \sqrt{2}r_s~,
\end{equation}
where $v_F$ is the Fermi velocity, $E$ is the magnitude of the electric field in the $z$ direction,
 $a_B$ is the Bohr radius, $P$ is the matrix element of the momentum operator between
 the conduction and the valence band at the zone center, $E_g$ is the gap,
 $\Delta_{so}$ is the SO splitting of the lowest hole band,
 and $r_s$ is the average distance between the electrons in units of $a_B$.

Notice that we have already assumed that the direction of the
external magnetic field coincides with the direction  of the
equilibrium magnetization ($\e3$, by definition). This is not
generally true in the presence of SO coupling (except for special
cases such as $\Theta =0$), but the angle between $\Hv_0$ and
$\Mv_0$ is of order $\left(\frac{\hbar \alpha k_F}{E_F}\right)^2$
and will be neglected henceforth. The magnitude of the
magnetization,  neglecting Coulomb interactions, is simply given
by $M_0 = \left(\frac{g \hbar}{2}\right)^2 N(0)H_0$ where $N(0) =
\frac{m}{\pi \hbar^2}$ is the free-particle density of states in
two dimensions.

The SO torque, defined in Eq.~(\ref{spintorque}) is straightforwardly calculated to be
\begin{equation} \label{spintorque2}
\mbox{\tauv} = \frac{\alpha}{\hbar}\sum_n \left[\ez \left(\hat \pv_n \cdot \Sv_n\right)
   -\hat \pv_n \hat S_{nz}\right]~.
\end{equation}
In order to facilitate the calculations in the limit of weak
SO coupling it is convenient to express the torque in
terms of the components $p_x$ and $p_y$ of the momentum in the
plane, as well as the components $S_1$, $S_2$ and $S_3$ of the
spin in the coordinate system shown in Fig.~1.  This is
advantageous when the only significant anisotropy in spin space is
caused by the magnetic field in the $\e3$ direction.  The result
of this rewriting is
\begin{equation} \label{spintorque3}
\left\{
\begin{array}{l}
\hat \tau_1 = \alpha\hbar^{-1}\sum_n \left(\hat S_{n3}\hat p_{nx} - \hat S_{n2}\hat p_{ny} \sin \Theta   \right)\\
\hat \tau_2 =\alpha\hbar^{-1}\sum_n \left(\hat S_{n3}\hat p_{ny}\cos \Theta + \hat S_{n1}\hat p_{ny} \sin \Theta   \right)\\
\hat \tau_3 =\alpha\hbar^{-1}\sum_n \left(-\hat S_{n1}\hat p_{nx}- \hat S_{n2}\hat p_{ny} \cos \Theta   \right)
\end{array}
\right.
\end{equation}
Notice that these equations do not involve $p_z$ --  not just because our model system is two-dimensional but, more fundamentally, because a motion along the direction of the electric field does not produce spin-orbit coupling.

The commutators $[\hat \tau_i,\hat S_j]$ that appear in Eq.~(\ref{Gamma})  can also be straightforwardly calculated:
\begin{eqnarray} \label{commutators}
-i [\hat \tau_1,\hat S_1] &=&\alpha \sum_{n}\left(\hat S_{n3}\hat p_{ny}\sin\Theta+\hat S_{2n} \hat p_{nx}\right)\nonumber \\
-i [\hat \tau_2,\hat S_2] &=&-\alpha \sum_{n}\left(\hat S_{n1}\hat p_{ny}\cos \Theta-\hat S_{n3} \hat p_{ny} \sin\Theta \right)\nonumber\\
 -i [\hat \tau_1,\hat S_2] &=&-\alpha \sum_{n}\hat S_{n1}\hat p_{nx} \nonumber \\
 -i [\hat \tau_2,\hat S_1] &=&\alpha \sum_{n}\hat S_{n2}\hat p_{ny}\cos\Theta~.
\end{eqnarray}
Their expectation values in the ground state are of second order in $\alpha$
and are closely related the change in energy of the electron gas due to the SO interaction
 (i.e. they control the magnetic anisotropy induced by the SO interaction).

The fact that in this model the torque is linear in both spin and momentum, i.e. proportional to
 the spin-current, allows us to establish an exact connection between Gilbert damping
  and spin-channel conductivity, defined in terms of spin-current--spin-current response functions.
  Unfortunately, this relation does not hold in more realistic models of the SO coupling,
  in which the electric field that is responsible for the SO coupling depends on position rather
   than being constant.  But we will argue in the concluding section that a similar relation,
   involving a spatially varying SO coupling constant and a local spin conductivity can be justified
   if certain conditions are met.

\subsection{Isotropic case}\label{IsotropicCase}
\subsubsection{Non-interacting 2DEG}

Let us begin with the simplest case in which the magnetic
 field and the magnetization  are parallel to the $z$ axis ($\Theta = 0$).
 The system is invariant for rotations about the $z$ axis, which coincides with the $3$ axis, so we have
\begin{eqnarray}\label{tau11}
\langle \langle\hat \tau_1;\hat \tau_1\rangle\rangle_\omega &=&
\langle \langle\hat \tau_2;\hat \tau_2\rangle\rangle_\omega
\nonumber\\&=& \frac{\alpha^2}{\hbar^2}\langle \langle\sum_n \hat S_{n3}\hat p_{nx};
\sum_n \hat S_{n3}\hat p_{nx}\rangle\rangle_\omega~.
\end{eqnarray}

Let us introduce the {\it longitudinal spin-channel conductivity},
$\sigma_{s\parallel}$, as the constant of proportionality  between
the
 spin-current $j_\up - j_\down$ and an electric field $E_s$, which
 acts with opposite signs on the spin up and the spin down components of the electron liquid.
  \footnote{Notice that such a field creates, in general, also a charge current.
  However, only the spin current is involved in the definition of $\sigma_{s\parallel}$.}
   $\up$ and $\down$ denote the directions parallel and antiparallel
   to the magnetic field (the $\e3$ axis).  $\sigma_{s\parallel}$ is
   related to the spin-current spin-current response
function by the Kubo formula
\begin{equation}\label{sigmasparallel}
\Re e \sigma_{s\parallel} (\omega) = - \frac{4e^2}{m^2 {\cal V}}
\Im m \frac{\langle \langle\sum_n \hat S_{n3}\hat p_{nx};\sum_n
\hat S_{n3}\hat
p_{nx}\rangle\rangle_\omega}{\hbar^2\omega}~.\end{equation} We now
discern a simple relationship between the imaginary part of the
torque-torque response function and the spin-channel conductivity,
namely
\begin{equation}\label{Gamma&Sigma}
\left.\frac{1}{{\cal V} }\frac{\Im m \Gammat (\omega)}{\omega}\right\vert_{sym} =
-\frac{m^2\alpha^2}{4 e^2}\Re e\sigma_{s\parallel}(\omega) \onet
\end{equation}

Substituting this in Eqs.~(\ref{Gilbert11}--\ref{Gilbert22})
 we arrive at the following expression for the Gilbert damping constant:
\begin{equation}\label{lambda-sigmas}
\lambda= \lambda_1 = \lambda_2 = \frac{m^2}{e^2}\frac{g
\alpha^2}{4 M_0}\Re e\sigma_{s\parallel}(0)~,
\end{equation}
Let us emphasize that the relation between equilibrium
magnetization and external field in this model is not affected by
SO interactions as long as both spin-orbit split bands are
occupied. The e-e interaction, on the other hand, would reduce the
value of $\omega_0$ required to produce a given $M_0$, since part
of the magnetization would then arise from the exchange
interaction.   However what will be needed in the subsequent
calculations is not $\omega_0$ per se, but the Zeeman splitting
that it produces on single particle energy levels.  The latter is
brought back to the non-interacting value once exchange is
included in the single particle energy levels.  So it seems
permissible to ignore the e-e correction to $\omega_0$, as long as
we do not include e-e effects in the spectrum of single particle
excitations.
%{\bf \color{red} Presence of
%$e-e$ does not change magnetization $M_0$.  The reason is that the
%Zeeman splitting depends on $M_0$, so even if the effective field
%consists of the two contributions (exchange and external one) the
%total Zeeman splitting does not change. Further we believe that
%single particle properties like  the electron-hole excitations
%will not be much different when the exchange is present. As a
%result the calculations of Gilbert damping using unchanged $M_0$
%should be justified.}
Finally, it can be shown that the spin-orbit anisotropy
stiffness $\ST_{so}$ for this model begins with terms of order
$(\alpha/v_F)^4$ and is therefore negligible for typical values
of $\alpha$ and $v_F$.

Having considered all this we can express $\lambda$ in the elegant form
\begin{equation}\label{lambda.elegant}
\lambda=  \frac{\bar \alpha^2 E_F\tau}{\hbar p} \frac{\Re e\sigma_{s\parallel}(0)}{\sigma_D}~,
\end{equation}
where $\bar \alpha$ is the Rashba velocity in units of the Fermi velocity
\begin{equation}
\bar \alpha \equiv \frac{\alpha}{v_F}~,
\end{equation}
$\sigma_D$ is the usual Drude conductivity
\begin{equation}
\sigma_D = \frac{n e^2 \tau}{m}~,
\end{equation}
$\tau$ is the transport scattering time, $E_F$ is the Fermi energy, and $p$ is the degree of polarization
of the electron gas
\begin{equation}
p= \frac{\hbar \omega_0}{2 E_F} = \frac{2M_0}{g \hbar n}~.
\end{equation}

The next task is the calculation of the spin-channel conductivity.  In the absence of  Coulomb interaction
 this can be done even without assuming that  $\bar \alpha$ is  small compared to $p$ and $\hbar/E_F\tau$.
  We supply the details of the calculation in Appendix A.  The result is
\begin{equation}\label{SpinChannelConductivity}
\Re e \sigma_{s\parallel}(0) = \sigma_D \cos^2\delta \frac{1+
\frac{1}{2}(\Omega\tau)^2(1+\cos^2\delta)}
{\cos^2\delta+\frac{1}4{}(\Omega\tau)^2(1+\cos^2\delta)^2}~,
\end{equation}
where we have defined
\begin{equation}
\Omega\equiv \sqrt{\omega_0^2+\alpha^2k_F^2}
\end{equation}
and
\begin{equation}\label{cosdelta}
\cos \delta \equiv \frac{\omega_0}{\Omega} = \frac{p}{\sqrt{p^2+\bar\alpha^2}}~,
\end{equation}
and correspondingly:
\begin{equation}\label{sindelta}
\sin \delta \equiv \frac{\alpha k_F}{\Omega} =
\frac{\bar\alpha}{\sqrt{p^2+\bar\alpha^2}}~.
\end{equation}
We have assumed, for simplicity, that the transport scattering
time $\tau$ is the same for up- and down-spin electrons.  Notice
that, in the absence of SO coupling, one recovers
$\sigma_{s\parallel}(0)=\sigma_D$
 as expected since up- and down-spin components are completely decoupled and have the same mobility.
 If, on the other hand,  the external field frequency $\omega_0$ is set to zero when $\alpha$ is still finite
 we get $\sigma_{s\parallel}(0) = 0$.  Again, this is not surprising in view of the fact that
   the SO interaction causes a steady precession of the spin in a
plane perpendicular to $\pv$, effectively suppressing the average
$z$-component of the spin. Putting
Eq.~(\ref{SpinChannelConductivity}) back into
Eq.~(\ref{lambda.elegant}) we arrive at
\begin{equation}\label{lambdani}
\lambda_{ni}= p\frac{ E_F\tau\sin^2\delta}{\hbar}\frac{1+
\frac{1}{2}(\Omega\tau)^2(1+\cos^2\delta)}{\cos^2\delta+\frac{1}{4}(\Omega\tau)^2(1+\cos^2\delta)^2}~,
\end{equation}
where the subscript $ni$ stands for {\it non-interacting}.  Notice that this expression
 vanishes  in the limit of zero magnetization, i.e. for $p \to 0$.

\begin{figure}[thb]
\vskip 0.27 in
\includegraphics[width=3.5in]{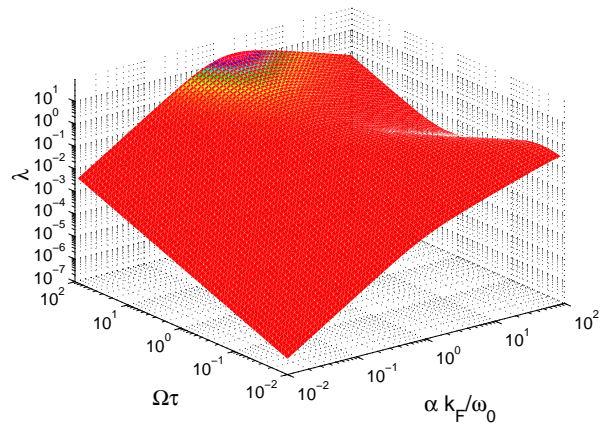}
\caption{Gilbert damping constant as a function of SO to magnetic
field ratio $\alpha k_F/\omega_0$ and  disorder scale $\Omega\tau$.  We assumed
$n_{2D}=10^{12}cm^{-2}$ and the polarization p=0.1}
\end{figure}

\begin{figure}[thb]
\vskip 0.27 in
\includegraphics[width=3.5in]{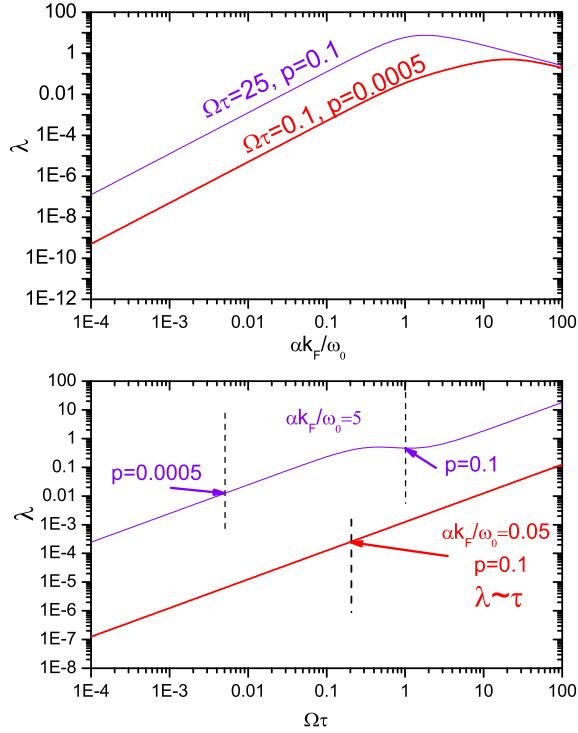}
\caption{(a) The Gilbert damping constant as a function of SO to magnetic
field ratio for two different values of disorder scale
$\Omega\tau$ corresponding to the clean and dirty limits. The
magnetic field and as a consequence the polarization is fixed,
while the strength of spin-orbit interactions and $\tau$ changes
so $\Omega\tau=$ const. The curves for $p=0.1$ and $p=0.0005$ and
the ratio $\Omega\tau = 0.1$ are indistinguishable. (b) The
Gilbert damping as a function of disorder scale for two different
values of $\alpha k_F/\omega_0$ corresponding to the strong and
weak SO couplings. The magnetic field and so the polarization and
the spin orbit coupling strength are fixed. The condition of
validity of our calculations $E_F\tau \gg 1$ is on the right to
the dashed line. The curves for $p=0.1$ and $p=0.0005$ and the
ratio $\alpha k_F/\omega_0 = 5$ are indistinguishable. }
\end{figure}

The Gilbert damping constant,  as given by Eq.~(\ref{lambdani}) is
plotted in Fig.~2 as a function of two parameters: $\Omega\tau$ --
measuring the effectiveness of spin precession during an elastic
mean free path $\tau$ --  and $\alpha k_F/\omega_0 = \bar
\alpha/p$  the ratio of the SO effective magnetic field to the
external magnetic field.  Clearly $\Omega\tau \gg 1$ corresponds to the
``clean" limit and $\Omega\tau \ll 1$ to the ``dirty" limit.

Fig.~3 shows the two-dimensional cross-sections of Fig.~2.  In  Fig.~3a we see the Gilbert constant
 plotted as a function of the spin-orbit to polarization ratio for two values of $\Omega\tau$, one
 in the clean limit and one in the dirty limit.   In the weak SO coupling regime ($\alpha k_F/\omega_0 \ll 1$)
   the Gilbert constant grows as $(\alpha k_F/\omega_0)^2$,  while for strong SO coupling it saturates and
begins to decrease linearly beyond a certain value $\alpha k_F/\omega_0$.
The quadratic increase of $\lambda$ for small $\alpha$ is easily accounted for by
the growth of the $\sin^2 \delta$ factor in  Eq.~(\ref{lambdani}) --  see Eq.~(\ref{sindelta}).
For large $\alpha$, on the other hand, $\sin \delta$ approaches the maximum value of $1$
 and $\lambda$ becomes essentially proportional to $\tau$.  The decrease in $\lambda$ in this regime reflects
 the decrease in $\tau$ along a curve on which  $\Omega$ grows  while $\Omega \tau$ remains constant.

Fig.~3b shows
the Gilbert damping as a function of $\Omega\tau$ for two
different values of  $\alpha k_F/\omega_0$. For
weak SO coupling (lower solid line) the Gilbert damping constant is proportional to $\tau$
in both the clean and dirty limits. This is not surprising given that in this regime the Gilbert damping is
essentially proportional to the Drude conductivity.  For  strong SO coupling,  however, we observe
an interesting non-monotonic behavior of $\lambda$ (upper solid curve) in the transition region
 between the dirty regime ($\Omega \tau \ll 1$)  and the clean one $\Omega \tau \gg 1$.
  This region is defined by the inequality  $2\cos \delta < \Omega \tau < \sqrt{2}$,
  and obviously appears only if $\cos \delta$ is sufficiently small, i.e., if the SO coupling is sufficiently strong.

There is a limit, however, on how large the spin-orbit coupling can be made for a given value of $\Omega \tau$. Recall that our treatment of disorder is justified for $E_F\tau \gg 1$ and becomes uncontrolled when this inequality is violated.  Since $E_F\tau = \frac{\Omega \tau}{2p\sqrt{1+(\alpha k_F/\omega_0)^2}}$ we see that  $\Omega \tau/2p$  must be larger than $\frac{\alpha k_F}{\omega_0}$  or, equivalently, $\Omega \tau > 2 \bar \alpha$ if $E_F\tau \gg 1$ is to be satisfied.  In Fig. 3b the regions of validity of this condition, for three different values of $\bar \alpha$,  extend to the right of the three vertical dashed lines.  We see that the non-monotonic behavior for larger $\bar \alpha$  occurs well within the region in which our treatment of disorder is justified and can be observed if $\bar \alpha$ is much larger than $p$, so that $\cos \delta$ is small.

%
%This non-monotonic behavior of Gilbert damping is a
%consequence of the competing scales: magnetic field, spin-orbit
%interactions and disorder and will appear for $\Omega\tau$ in a
%range: $\cos\delta \ll \Omega\tau \ll 1$. However, our
%calculations are valid for $E_F\tau >1$, which is to the right of
%the dashed lines in Fig.~3b. Therefore the nonmonotonic behavior
%of Gilbert damping can be observed for strong spin-orbit couplings
%and very small polarizations.

In a naive microscopic picture the Gilbert damping could be obtained by substituting the D'yakonov-Perel spin relaxation rate\cite{Halperin04}
\begin{equation}\label{Dyakonov}
\frac{1}{\tau_{s}} =\frac{2\tau(\alpha k_F)^2}{1+(2\tau\alpha k_F)^2}
\end{equation}
into the phenomenological Bloch equation for spin dynamics.  This results in the formula
\begin{equation}\label{lamb_exchange}
\lambda \sim \frac{\tau_{s}}{1+(\omega_0\tau_{s})^2}\,.
\end{equation}
which reproduces Eq.~(\ref{lambdani}) either in the weak-SO limit with $\alpha k_F\ll1/\tau,\omega_0$ or in the absence of the applied field, $\omega_0=0$, where Eq.~(\ref{Dyakonov}) becomes the correct Bloch spin relaxation rate of the unpolarized Rashba 2DEG \cite{Halperin04}.  We note, however, that the full structure of Eq.~(\ref{lambdani})  cannot be completely captured by the
D'yakonov-Perel spin relaxation rate (\ref{Dyakonov}). In particular in the clean limit the combined Eqs.~(\ref{Dyakonov}) -- (\ref{lamb_exchange}) would lead to the scaling
of $\lambda$ as $1/\tau$, which tends to zero, while our Eq.~(\ref{lambdani}) gives
scaling of $\lambda$ as $\tau$, which tends to infinity.

\begin{figure}[thb]
\vskip 0.27 in
\includegraphics[width=3.5in]{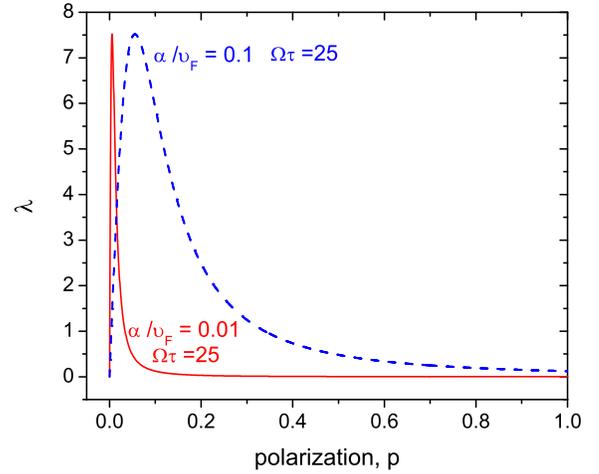}
\caption{The Gilbert damping constant as a function of polarization for two
different values of renormalized spin-orbit coupling
$\bar{\alpha}=\alpha/\upsilon_F$.}
\end{figure}

Fig.~4 shows the Gilbert damping as a function of polarization for
two different values of spin-orbit coupling. One can see that
Gilbert damping increases with $p$ for $p<\bar{\alpha}$ and
decreases with further increase of $p$ above $\bar{\alpha}$. Tunable magnetization
damping is very desirable since it allows to reduce post-switching
magnetization precession \cite{Ingvarsson04}. Fig.~4 shows the
possibility of tuning of the Gilbert damping by changing
temperature in a polarized 2DEG with spin-orbit interactions.

The $\omega$ dependent  spin-channel conductivity, can be obtained
through the replacement of $\frac{1}{\tau}$ by $\frac{1}{\tau} - i
\omega$ (see also appendix A) and has a form:
 \begin{eqnarray}\label{SpinChannelConductivity.omega}
&&\Re e \sigma_{s\parallel}(\omega) = \sigma_D
\frac{\cos^2\delta}{1+(\omega\tau)^2} \nonumber \\
&\times&\frac{1+ (\Omega\tau)^2[\frac{1}{2}\sin^2\delta+(
\cos\delta+\omega/\Omega)^2]}
{\cos^2\delta+(\Omega\tau)^2[\frac{1}{2}\sin^2\delta+\cos\delta(
\cos\delta+\omega/\Omega)]^2}~,\nonumber\\
\end{eqnarray}
which gives
\begin{eqnarray}\label{lambdani.omega}
&&\lambda_{ni}(\omega)= p\frac{\bar \alpha^2 E_F\tau}{\hbar
(p^2+\bar\alpha^2)} \frac{1}{1+(\omega\tau)^2}\nonumber\\ &\times&
\frac{1+ (\Omega\tau)^2[\frac{1}{2}\sin^2\delta+(
\cos\delta+\omega/\Omega)^2]}{\cos^2\delta+(\Omega\tau)^2[\frac{1}{2}\sin^2\delta+\cos\delta(
\cos\delta+\omega/\Omega)]^2}~.\nonumber\\
\end{eqnarray}
These equations reduce to Eqs.~(\ref{SpinChannelConductivity})
and~(\ref{lambdani}) respectively if we set $\omega=0$.

 Notice the very different limiting behavior of $\lambda$ as a
function of
 the disorder strength for $\omega=0$ and $\omega$ finite. In the first case,
 $\lambda$ tends to infinity as the scattering time $\tau$ increases (i.e. for decreasing disorder).
 At finite $\omega$ however, the Gilbert constant goes to zero for $\tau$ tending to infinity.
This is the proper ``clean limit". In the opposite limit of strong
disorder, the Gilbert damping decreases monotonically to zero.
This is a manifestation of the D'yakonov-Perel effect: in a
strongly disordered system the instantaneous axis of
spin-precession changes too rapidly to allow an effective loss of
spin orientation.

Instead of using the equation of motion to calculate the Gilbert
damping, we could evaluate directly the imaginary  part of the
transverse spin response function. The two approaches are equivalent
and the details of the calculations of the transverse spin response
function are summarized in Appendix~D. We will see in the following, however,
that the treatment in terms of the spin-channel conductivities allow us to gain
useful intuition for the inclusion of $e-e$ interaction effects.

\subsubsection{Interacting 2DEG}

The role of the Coulomb interaction in spin transport was analyzed
in detail in Refs.~\onlinecite{Amico00,Amico02}, but only in the
absence of SO coupling.  In the spin-polarized electron liquid,
with identical scattering times for up and down spins and
$\alpha=0$ the result is
\begin{equation}\label{Drude&Coulomb}
\Re e \sigma_{s\parallel}(0) = \sigma_D \frac{1+p^2\gamma\tau}{1+\gamma \tau}~,
\end{equation}
where $\gamma$ is the spin-drag coefficient -- the rate of
momentum transfer due to the Coulomb interaction between up and
down spins.  The derivation of Eq.~(\ref{Drude&Coulomb}) is
presented for completeness in Appendix B.

Notice that the effect of the spin Coulomb drag vanishes (as expected)
 when the electron gas is fully spin polarized, i.e. when $p=1$.
 This is because in this limit there are no minority spin carriers to
 exert a drag on the majority spin carriers.
  More generally, the form of Eq.~(\ref{Drude&Coulomb}) can be understood in terms of
  the coupling between charge and spin currents in a spin-polarized electron gas.
   A ``spin electric field"   $E_s=E_\uparrow=-E_\downarrow$  drives not only a spin current,
   but also a charge current $j_c =
 p \sigma_D  E_s$ (notice that $\gamma$ does not enter here).
  The charge-current acts on the spin current as an additional spin electric field
  $ \frac{m \gamma}{ne^2} p j_c = \gamma \tau p^2 E_s$.
  The spin current then responds in the usual way (i.e. with the  unpolarized spin conductivity
  $\sigma_s\vert_{p=0} = \frac{\sigma_D}{1 +\gamma\tau}$) to an enhanced spin electric field
  $E_s (1 + \gamma p^2 \tau)$.  This gives Eq.~(\ref{Drude&Coulomb}).

 The low-temperature behavior of $\gamma$ is approximately given by the
formula
\begin{eqnarray}\label{gamma2}
\gamma(T) &=& \frac{E_F}{\hbar}\left(\frac{k_BT}{E_F}\right)^2
\frac{\pi r_s^2}{6(1-p^2)^{3/2}}\nonumber\\&\times&
\int_{0}^{1} \frac{dx}{{(\sqrt{x}+\frac{r_s}{\sqrt{2(1-p)}})^2}\sqrt{1-x}\sqrt{1-a x}}~, \nonumber\\
\end{eqnarray}
where $a=(1-p)/(1+p)$.\footnote{We take this opportunity to correct an error in
Eq.~15 in Ref.~[35].  $x$ should be replaced by  $x^2$ and $\alpha x$ by $(\alpha
x)^2$ under the square roots in the denominator on the r.h.s of that equation.} For zero
polarization, $\gamma$ diverges logarithmically as can be easily
by putting $p=0$ under integral in Eq.~(\ref{gamma2}). Thus
$\gamma$  scales quadratically with temperature for $T/T_F\ll
1$, but has a non-monotonic behavior as a function of
polarization.

\begin{figure}[thb]
\vskip 0.27 in
\includegraphics[width=3.5in]{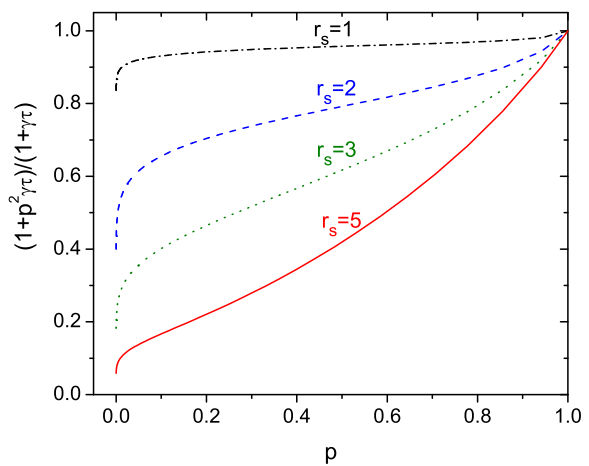}
\caption{The behavior of renormalization factor $(1 +p^2 \gamma
\tau)/(1+\gamma \tau)$ associated with electron-electron
interactions as a function of polarization for various $r_s$.}
\end{figure}

  The behavior of the spin conductivity renormalization factor $(1 +p^2 \gamma \tau)/(1+\gamma \tau)$
  of Eq.~(\ref{Drude&Coulomb})  is shown in Fig.~5
  as a function of $p$ for several values of $r_s$,  assuming $E_F \tau/\hbar =10/r_s^2$ and $k_BT/E_F = r_s^2/10$.
  We see that the reduction of the spin conductivity  and hence of $\lambda$ due to Coulomb interaction is
   significant for small and intermediate polarizations, especially at larger values of $r_s$.

The calculation of the spin drag coefficient in the presence of SO
coupling poses, of course, a more difficult problem. However, in
view of the fact that the SO energy scale is
 usually much smaller than the Coulomb energy scale such a detailed calculation is not urgently needed.
 Coulomb interaction corrections are adequately taken into account by multiplying the non-interacting result of
Eq.~(\ref{lambdani}) by the correction factor on the right hand
side of E.~(\ref{Drude&Coulomb}).  This gives
\begin{equation}\label{lambdafull}
\lambda = \lambda_{ni}\frac{1+p^2\gamma\tau}{1+\gamma \tau}~.
\end{equation}
Eq.~(\ref{lambdafull})  has the interesting property that it scales as $\tau$ for
$\gamma p^2\tau \gg 1$ and $ \gamma \tau\gg 1$ as well as in opposite limit when
 $\gamma p^2\tau\ll 1$ and $ \gamma \tau \ll 1$. The $e-e$ interactions plays
 a role mostly in the regime $1 \ll \gamma\tau \ll 1/p^2$ which shrinks to zero as $p$ increases.
The behavior of Gilbert constant as a function of $\gamma\tau$ and
polarization $p$  is presented in Fig.~6 and Fig.~7. For strong
polarization the factor $(1+p^2\gamma\tau)/(1+\gamma\tau)\sim 1$
and the Gilbert damping scales with a scattering time $\tau$. The
situation looks different for a very weak polarization (see
solid line in Fig.~7):  the Gilbert damping scales as
$\tau/(1+\gamma\tau)$ for a weak $e-e$ interactions but for strong
$e-e$ interactions it saturates to a constant value proportional to $1/\gamma$.

The dependence of  $\lambda$  on polarization is more complicated. This is a
consequence of the fact that the spin-Coulomb drag depends on
polarization. For weak $e-e$ interaction $\lambda$ increases linearly with $p$ for small
$p$ and saturates for large $p$. For strong $e-e$
interactions ($\gamma\tau >1$)  $\lambda$  increases
linearly with $p$ for small $p$, and increases even faster
than $p$ for strong polarization.
\begin{figure}[thb]
\vskip 0.27 in
\includegraphics[width=3.5in]{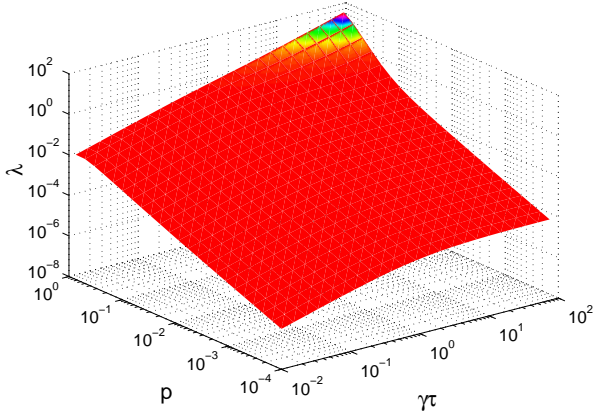}
\caption{Gilbert damping constant as a function of $\gamma\tau$ and
polarizatioon p for $n_{2D}=10^{12}cm^{-2}$ and
$m_{eff}=0.067m_e$. We chose the limit of weak spin orbit coupling
$\alpha k_F/\omega_0 =0.02$ and $T/T_F =0.02$.}
\end{figure}
\begin{figure}[thb]
\vskip 0.27 in
\includegraphics[width=3.5in]{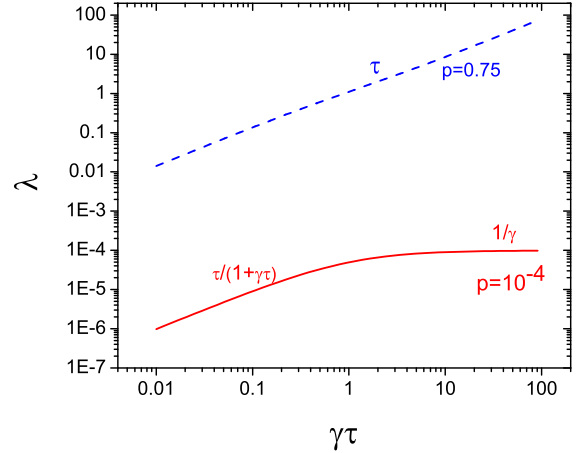}
\caption{The Gilbert damping constant as a function $\gamma\tau$, the
two-dimensional cross-section through the 3D plot presented in
Fig.5.}
\end{figure}

It is also possible to include  the frequency dependence of the
spin Coulomb drag correction through the replacement of
$\frac{1}{\tau}$ by $\frac{1}{\tau} - i \omega$. A straightforward
calculation gives
\begin{equation}\label{Gilbertdamping.omega}
\lambda (\omega)= \lambda_{ni}(\omega)\frac{(1+\gamma\tau)(1+p^2\gamma\tau)+(\omega\tau)^2}
{(1+\gamma\tau)^2+(\omega\tau)^2}~.
\end{equation}

\subsection{Anisotropic case}
\label{AnisotropicCase}

Let us now consider the general anisotropic case, in which the direction $\e3$ of the magnetization forms an angle $\Theta$ with the direction $\ez$ of the electric field that generates the SO effective magnetic field.  An exact calculation of the torque-torque response function is complicated, even in the noninteracting case, by the lack of rotational symmetry about the $\e3$ axis.  However, in the limit of small $\alpha$,  the spin degrees of freedom are decoupled from the orbital degrees of freedom and we can take advantage of rotational symmetry about the $\e3$ axis in spin space to simplify the calculation.    What happens is that response functions such as $\langle \langle\sum_n \hat S_{n3}\hat p_{nx};\sum_n \hat S_{n\perp}\hat p_{nx}\rangle\rangle_\omega$ vanish by symmetry  when
$\hat S_\perp$ denotes one of the two transverse components, $\hat S_1$ or $\hat S_2$, of the spin operator.   On the other hand, the response function $\langle \langle\sum_n \hat S_{n\perp}\hat p_{nx};\sum_n \hat S_{n\perp}\hat p_{nx}\rangle\rangle_\omega$ does not depend on which component  $\hat S_{n\perp}$ we choose to consider.

Going back to Eqs.~(\ref{Gilbert11}-\ref{Gilbert22}) and making use of Eq.~(\ref{spintorque3}) we see that,  to second order in $\bar \alpha$ we have
\begin{equation}\label{lambda1-anisotropic}
\lambda_1=  \frac{\bar \alpha^2 E_F\tau}{\hbar p}
\left[\frac{\Re e\sigma_{s\parallel}(0)}{\sigma_D}+\sin^2 \Theta\frac{\Re e\sigma_{s\perp}(0)}{\sigma_D} \right]~,
\end{equation}
and
\begin{equation}\label{lambda2-anisotropic}
\lambda_2=  \frac{\bar \alpha^2 E_F\tau}{\hbar p} \left[\cos^2
\Theta\frac{\Re e\sigma_{s\parallel}(0)}{\sigma_D}+\sin^2
\Theta\frac{\Re e\sigma_{s\perp}(0)}{\sigma_D} \right]~,
\end{equation}
where  the {\it transverse spin-channel conductivity} $\sigma_{s\perp}$ is defined,
 in analogy to Eq.~(\ref{sigmasparallel})  as
\begin{equation}\label{sigmasperp}
\Re e\sigma_{s\perp}(\omega)=  - \frac{4e^2}{m^2 {\cal V}} \Im m
\frac{\langle \langle\sum_n \hat S_{n\perp}\hat p_{nx};\sum_n
\hat S_{n\perp}\hat p_{nx}\rangle\rangle_\omega}{\hbar^2 \omega}~.
\end{equation}

The calculation of  $\sigma_{s\perp}$ in the absence of SO coupling {\it and}
 $e-e$ interactions is straightforward and yields
\begin{equation}
\Re e\sigma_{s\perp}(0)=\frac{\sigma_D}{1+(\omega_0\tau)^2}
\end{equation}
Notice that this differs from the longitudinal spin-channel
conductivity $\sigma_{s\parallel}$, calculated in the same
approximation, simply by the factor  $[1+(\omega_0\tau)^2]^{-1}$.
 This takes into account the non-conservation (precession) of the transverse spin caused by
 the magnetic field in the $\e3$ direction.  Therefore, in the limit of weak SO coupling
  and no interaction we get
\begin{equation}\label{lambda1.aniso}
\lambda_1=  \frac{\bar \alpha^2 E_F\tau}{\hbar p}
\left[1+\frac{\sin^2 \Theta}{1+(\omega_0\tau)^2} \right]~,
\end{equation}
and
\begin{equation}\label{lambda2.aniso}
\lambda_2=  \frac{\bar \alpha^2 E_F\tau}{\hbar p} \left[\cos^2
\Theta+\frac{\sin^2 \Theta}{1+(\omega_0\tau)^2} \right]~.
\end{equation}

The effect of the spin Coulomb drag can be included, at the same level of approximation,
 by multiplying the two spin-channel conductivities by appropriate renormalization factors.
 Namely, we use Eq.~(\ref{Drude&Coulomb}) for  $\sigma_{s\parallel}(0)$, and
\begin{equation}\label{sigmaperpCoulomb}
\Re e  \sigma_{s\bot}(0) = \sigma_D \frac{1+\gamma_{\bot}\tau}
{(\omega_0\tau)^2+(1+\gamma_{\bot}\tau)^2}~,
\end{equation}
 where $\gamma_\perp$ is a transverse-spin analogue of the longitudinal
 spin Coulomb drag coefficient, as discussed in Appendix C.
%  (The somewhat simpler form of this transverse Coulomb renormalization,
%  compared to that of Eq.~(\ref{Drude&Coulomb}) reflects the fact that the transverse spin channel,
%  at $\alpha=0$ is decoupled from the density and the longitudinal spin
%  channel).

An unsatisfactory feature of  Eqs.~(\ref{lambda1.aniso})
and ~(\ref{lambda2.aniso}) is that they are valid only for $\alpha k_F \ll \omega_0$.
Notice however that the angular dependence of the Gilbert constants in these equations reduces
 to simple angular factors multiplying the isotropic result.  This suggests that we take care of the
  problem simply by multiplying the full isotropic non-interacting result for $\lambda(\omega)$,
  given by Eq.~(\ref{lambdani}), by the same angular factor.  The resulting
  expressions:
\begin{equation}\label{lambda1.final}
\lambda_1=  \lambda_{ni}\left[\frac{\Re e\sigma_{s\parallel}(0)}{\sigma_D}+
\sin^2 \Theta\frac{\Re e\sigma_{s\perp}(0)}{\sigma_D} \right]~,
\end{equation}
and
\begin{equation}\label{lambda2.final}
\lambda_2=  \lambda_{ni}\left[\cos^2 \Theta\frac{\Re e\sigma_{s\parallel}(0)}
{\sigma_D}+\sin^2 \Theta\frac{\Re e\sigma_{s\perp}(0)}{\sigma_D} \right]~,
\end{equation}
with $\lambda_{ni}$, $\Re e\sigma_{s\parallel}(0)$ and $\Re e\sigma_{s\perp}(0)$ given
 by Eqs.~(\ref{lambdani}),
(\ref{Drude&Coulomb}) and~(\ref{sigmaperpCoulomb}) respectively, vanish for $\omega_0$ tending to zero,
and should be reasonable not only for  $\alpha k_F \ll \omega_0$ but also for  $\alpha k_F \gg \omega_0$
 provide both energies are small compared to the Fermi energy.
  Furthermore these formulas can be generalized to finite frequency simply replacing $\lambda_{ni}$ by
  $\lambda_{ni}(\omega)$ (Eq.~(\ref{lambdani.omega})) and replacing
  $\frac{1}{\tau}$ by $\frac{1}{\tau} - i \omega$ in the
   Coulomb renormalization factors.

\section{Generalizations and conclusion}
\label{Summary}

In this paper we have shown that (1) the Gilbert damping constant for a homogenously
magnetized electron gas with SO interactions can be exactly expressed
in terms of the torque-torque correlator which comes from SO
interactions;  (2) In the special case of the Rashba model the Gilbert damping can
be expressed in  terms of the spin-channel conductivity.  Based on this connection,
we have discussed the behavior of the Gilbert
damping constant in a two-dimensional electron gas with Rashba SO coupling as a function of magnetic field,
SO  interaction, $e-e$ interaction,   and disorder.   These calculations,
while based on linear response theory, do  nevertheless provide
the input for the nonlinear LLG equation, provided it is
recognized that $\lambdat$ depends on the instantaneous
orientation of the magnetization.

It should be clear that point (1) above is completely general, while point (2) depends on
 a specific feature of the Rashba spin-orbit interaction, namely, the fact that the electric field is
 independent of position.   This poses the question:  to what extent is the connection between Gilbert damping and spin conductivity transferrable to general, non-homogeneous, spin-orbit coupled systems?

In this conclusion we  argue that  the connection is indeed broadly applicable to itinerant-electron
ferromagnets, under assumptions similar to those that are used to justify the  local density approximation
(LDA) in the density functional theory of electronic structure.   We recall that  in a multi-band all-electrons
theory the SO interaction has the form:
\begin{equation}\label{generalSO}
\hat H_{SO} =-\frac{1}{2m^2 c^2}\sum_n \left(\hat \nabla V(\rv_n)\times \hat\pv_n
\right)\cdot \Sv_n~
\end{equation}
where $V(\rv)$ is the self-consistent Kohn-Sham potential, typically given by the local density approximation.
 The gradient of $V(\rv)$ defines a privileged direction at each point in space, which we denote $\ez(\rv)$.
 Then the spin-orbit interaction can be recast in the form

\begin{equation}\label{generalSO2}
\hat H_{SO} =-\sum_n \frac{\alpha(\rv_n)}{\hbar} \left[\ez (\rv_n) \times \hat \pv_n
\right]\cdot \Sv_n~,
\end{equation}
where $\alpha(\rv)$ is a position dependent SO
coupling constant, which reflects the local magnitude of the
electric field seen by the electron, and $\ez (\rv)$ is the local
direction of this field.  Another privileged direction is the direction of the local magnetization, which
 we call $\e3(\rv)$.  Thus we are back to our model hamiltonian~(\ref{modelHamiltonian})
  (we already remarked that the two-dimensionality of the model is not essential to our treatment)
with the crucial difference that $\alpha,\ez$, and $\e3$ are functions of position.
\footnote{Keeping the Coulomb potential on top of th Kohn-Sham single particle potential is no double counting,
 since the effect of the Coulomb interaction on the Gilbert damping, via the spin Coulomb drag,
 is of course not accounted for by the Kohn-Sham potential.}

Now in the spirit of the LDA let us assume that the Kohn-Sham potential, its gradient, and the
direction of the magnetization are all slowly varying on the appropriate microscopic length scale
(e.g., the electron-electron distance, or the size of the unit cell, or the scattering mean free path).  Then we can
equiparate each small volume element of the system to a homogeneously magnetized electron gas of density
$n$ and polarization $p$ pointing in the $\e3$ direction, with a SO coupling constant
 $\alpha(\rv)$ and a local anisotropy axis $\ez$.  With this identification, all the results obtained
 in the previous section become immediately applicable to each volume element.

 An obvious objection to this procedure is that the electronic density of real materials is not slowly
  varying on the atomic scale.  In spite of this, however, it is well known that the LDA works very well
  in materials because the exchange correlation potential is controlled by a spherical average of the exchange-correlation
  hole, which is reasonably close to the spherical hole of the homogeneous electron gas.
  We believe that a similar averaging may also work for the magnetization dynamics.
  Then the Gilbert damping would be given by the formulas derived in this paper,
   only with the appropriate coarse-grained values of $\alpha,\ez$, and $\e3$.

\section{Acknowledgements} We thank Allan H. MacDonald and Arne
Brataas for enlightening discussions. This work was supported by
NSF Grant No. DMR-0313681 and supported in part by the National
Science Foundation under Grant No. PHY99-07949.

\section{Appendix A: Longitudinal spin-channel conductivity for noninteracting electrons}
In the calculations that follow we set  $\hbar=1$ for convenience. To obtain the real part of  longitudinal
spin-channel conductivity we need to calculate $\Im m
\frac{\langle \langle\sum_n \hat S_{n3}\hat p_{nx};\sum_n \hat
S_{n3}\hat p_{nx}\rangle\rangle_\omega}{\omega}$. For a
magnetization parallel to the $z$ axis   (i.e. $\Theta=0$) the
above expression simplifies to
 $\Im m \frac{\langle\langle \hat S_z\hat P_x; \hat S_z\hat P_x\rangle\rangle_\omega}{\omega}$
where $\hat S_z$ is the  $z$ component of the spin operator and $\hat P_x$
is $x$ component of the total momentum. For non-zero
$\omega$ we have to evaluate the  following integral:
\begin{widetext}
\begin{equation}\label{general_formula}
\Im m \frac{\langle\langle
 \hat S_z \hat P_x; \hat S_z \hat P_x\rangle\rangle_\omega}{\omega{\cal V}}
=\frac{1}{\omega}\Im m \left\{\int \frac{d\epsilon}{2\pi i}
\int\frac{d^2p}{(2\pi)^2} p_x Tr[\hat S_z \hat
G(p,\epsilon+\omega) \hat\Lambda_x(\epsilon) \hat
G(p,\epsilon)]\right\}\nonumber\\
\end{equation}
where
\begin{equation}\label{G_p_omega}
G(p,\omega) =\frac{1}{\omega-\epsilon_p-\vec h_p\vec S
+i/2\tau_{\omega}}
\end{equation}
%and
%\begin{equation}\label{G_p_omega2}
%G(p,\hbar\omega+\epsilon)
%=\frac{1}{\hbar\omega+\epsilon-\epsilon_p-\vec h_p\vec S
%+i/2\tau_{\hbar\omega+\epsilon}}
%\end{equation}
is the disorder-averaged Green's function near the Fermi level,
$\epsilon_p$ is the kinetic energy relative to the Fermi level,
$\vec h_p =(\alpha p_y,-\alpha p_x,\omega_0)$ is the effective
magnetic field,  $\vec S =\vec \sigma/2$,where $\vec \sigma$ is
the vector of Pauli matrices and
\begin{equation}
\frac{1}{\tau_\omega} \equiv \frac{1}{\tau} {\rm sign}(\omega)~.
\end{equation}
\end{widetext}
Using the fact that the integration over energy involves only the
states around the Fermi energy, i.e. $\epsilon+\omega>E_F$ and
$\epsilon<E_F$ allows us to integrate over $\epsilon$ and cancel
out the $1/\omega$ on the r.h.s. of Eq.~(\ref{general_formula}).  Then one can see that the formula for a
non-zero $\omega$ can be obtained from the $\omega =0$ limit with
the substituition $1/\tau \to 1/\tau-i\omega$, leading to
Eq.~(\ref{lambdani.omega}).

 In the $\omega =0$ limit   $\hat G(\epsilon+\omega)$ and $\hat G (\omega)$ should be substituted by
$\hat G(0^+)$ and $\hat G(0^-)$, respectively. The ladder vertex
corrections are found by solving the self-consistent integral equation
for $\hat\Lambda$. For an electron-impurity potential of the form $U \delta(\rv)$ this equation is
\begin{equation}\label{vertex_func1}
 \vec \Lambda =\vec p S_z + U^2
\int\frac{d^2p'}{(2\pi)^2} \hat G(p',0^{+})\vec \Lambda \hat
G(p',0^{-})
\end{equation}
Its solution gives  the vertex correction of the following form:
\begin{equation}\label{vertex_func}
 \Lambda_x = p_x S_z + \frac{\alpha p_F^2}{2} \left(
 \frac{i\omega_0-1/\tau}{\Omega^2+\omega_0^2+2i\omega_0/\tau}S_{+} + h.c.\right)
\end{equation}
where $ S_{+}= S_x +i S_y$ and $U^2 = 1/m\tau$.

The calculation of the spin-channel
conductivity in Eq.~(\ref{general_formula})  without the vertex
corrections (a single bubble) gives:
\begin{eqnarray}\label{bubble}
\Im m \frac{\langle\langle \hat S_z \hat P_x; \hat S_z \hat
P_x\rangle\rangle_\omega}{\omega{\cal V}} = -\frac{mn\tau}{4}
\left(\frac{\Omega^2\tau^2cos\delta + 1}{\Omega^2\tau^2 +
1}\right)
\end{eqnarray}
while the final result including the ladder vertex corrections is:
\begin{eqnarray}\label{LSC}
\Im m \frac{\langle\langle \hat S_z \hat P_x; \hat S_z \hat
P_x\rangle\rangle_\omega}{\omega{\cal V}} = \frac{-mn\tau}{4}
\frac{\cos^2\delta\left[1+\frac{1}{2}\Omega^2\tau^2(1+\cos^2\delta)\right]}{\cos^2\delta
+\frac{1}{4}\Omega^2\tau^2(1+\cos^2\delta)^2}\nonumber\\
\end{eqnarray}
Using Eq.~(\ref{LSC}), and the definition of spin channel
conductivity we finally obtain the formula
Eq.~(\ref{SpinChannelConductivity}) for the real part of
longitudinal spin-channel conductivity:
$$\Re e
\sigma_{s\parallel}(0) = \sigma_D \cos^2\delta \frac{1+
\frac{1}{2}(\Omega\tau)^2(1+\cos^2\delta)}
{\cos^2\delta+\frac{1}{4}(\Omega\tau)^2(1+\cos^2\delta)^2}~$$

The single bubble calculation recovers the behavior of the
longitudinal spin-conductivity for weak SO interactions: Eqs.~(\ref{bubble}) and (\ref{LSC}) coincide for $\delta=0$, i.e. fora zero SO coupling. However, for
strong spin-orbit interactions the vertex corrections are
absolutely necessary. Moreover, only Eq.~(\ref{LSC}) gives the correct result
for zero magnetic field, i.e. $\Re e\sigma_{s\parallel}(0)=0$.

\section{Appendix B: Longitudinal spin-channel
conductivity in the presence of $e-e$ interactions}

In the absence of SO interactions the structure of the $2\times 2$
resistivity matrix  can be deduced from the equation of motion for
electrons  with spin $\sigma$ and velocity $\upsilon_{\sigma}$:
\begin{equation}\label{eq_motion}
m^*N_{\sigma}\vec\dot{\upsilon}_{\sigma} = -eN_{\sigma}\vec
E_{\sigma}+\vec F^C_{\sigma\bar{\sigma}}
-\frac{m^*}{\tau_{\sigma}}N_{\sigma}\vec\upsilon_{\sigma}
\end{equation}
where $m^*$ is the effective mass of the electrons, $ \bar{\sigma}
\equiv  -\sigma$,  $F^C_{\sigma\bar\sigma}$ is the average Coulomb force between electrons of opposite spin orientations, which is proportional
to the difference of their drift velocities. The Fourier transformation of
Eq.~(\ref{eq_motion}) leads to the following formula for the spin
current:
\begin{eqnarray}\label{spin_current}
i\omega\vec j_{\sigma} =\frac{-n_{\sigma}e^2}{m^*}\vec
E_{\sigma}(\omega) + \left(\frac{n_{\bar{\sigma}}}{n}\gamma
+\frac{1}{\tau_{\sigma}}\right)\vec j_{\sigma}(\omega)
-\frac{n_{\sigma}}{n}\gamma \vec
j_{\bar{\sigma}}(\omega)~.\nonumber\\
\end{eqnarray}
Using Eq.~(\ref{spin_current}), we can show that the resistivity
tensor has the form:
\begin{equation}\label{rho_tens} \rho =
\begin{pmatrix}
\frac{m^*}{n_{\uparrow}e^2}\left(-i\omega+\frac{1}{\tau_{\uparrow}}+\frac{n_{\downarrow}}{n}\gamma\right)
& -\frac{m^*}{ne^2}\gamma
\\-\frac{m^*}{ne^2}\gamma & \frac{m^*}{n_{\downarrow}e^2}
\left(-i\omega+\frac{1}{\tau_{\downarrow}}+\frac{n_{\uparrow}}{n}\gamma\right)
\end{pmatrix}
\end{equation}
Inverting the resistivity matrix we get the following formula for
the spin-channel conductivity $\sigma_{s{\|}}$:
\begin{equation}\label{spin_long}
\sigma_{s \|}(\omega) = \frac{\rho_{\downarrow
\downarrow}+\rho_{\uparrow \uparrow}+\rho_{\uparrow
\downarrow}+\rho_{\downarrow
\uparrow}}{\rho_{\uparrow\uparrow}\rho_{\downarrow\downarrow}-\rho_{\uparrow\downarrow}\rho_{\downarrow\uparrow}}
\end{equation}
Substituting the elements of the resistivity matrix into
Eq.~(\ref{spin_long}) one obtains:
\begin{equation}\label{Respin_long_omega}
\sigma_{s \|}(\omega) = \sigma_D\frac{[\omega^2+1/\tau^2(1+\gamma
\tau p^2)](\gamma +1/\tau) -\gamma
p^2\omega^2}{(\omega^2\tau+1/\tau)[\omega^2 + (1/\tau+\gamma)^2]}
\end{equation}
If $\omega =0$, Eq.~(\ref{Respin_long_omega}) simplifies  to
Eq.~(\ref{Drude&Coulomb}): $$\Re e \sigma_{s\parallel}(0) =
\sigma_D \frac{1+p^2\gamma\tau}{1+\gamma \tau}~,$$ For zero
polarization the spin and charge channels are decoupled and the
inverse of the effective scattering time consists of two
contributions, one connected with disorder and second one
associated with e-e interactions:
\begin{equation}\label{tau_dis_ee}
\tau^{-1}_{eff} = \tau^{-1} +\gamma
\end{equation}
For nonzero polarization the spin and charge channels are mixed
and the additional term $\gamma p^2\tau$ appears in the numerator.

\section{Appendix C: Transverse spin-channel
conductivity in the presence of $e-e$ interactions}

In this appendix we use the semi-classical approach (similar to
the one presented in a previous appendix) to calculate the real
part of the transverse spin-channel conductivity. According to
Eq.~(\ref{sigmasperp}): $$\Re e \sigma_{xx}=\Re e \sigma_{\bot}= -
\frac{4e^2}{m^2 {\cal V}} \Im m \frac{\langle \langle\sum_n \hat
S_{nx}\hat p_{nx};\sum_n \hat S_{nx}\hat
p_{nx}\rangle\rangle_\omega}{\hbar^2 \omega}$$ so we are looking
for a spin current response on the perturbation $p_xS_x$. This
perturbation is generated by the SU(2) vector potential $A_xS_x$ where
$S_x$ is the $x$-component of the spin operator. As a consequence one
can show that the ``electric field" that drives the change
of spin current does not depend on spin (see Eq.~(\ref{eq_motion_transverse1}) below). We neglect
SO interactions  in this calculation and work in the circularly polarized basis in which
the spin-conductivity is diagonal and has two components
$\sigma_s^{+}(\omega)$ and $\sigma_s^{-}(\omega)$.  The diagonal character of the conductivity in this basis is
equivalent to the statement that the current of right-handed circularly polarized electrons, $j_+$  does not interact with the current of left-handed circularly polarized
electrons  $j_{-}$.  The semiclassical equation of motion for these currents
is
\begin{equation}\label{eq_motion_transverse1}
\frac{e}{V}\sum_i \frac{d}{dt}(S_{i\pm }{\upsilon}_{ix}) =
-\sum_{i}\frac{e^2E}{m^*}- \sum_i\frac{eS_{i\pm
}\upsilon_{ix}}{\tau}-\sum_i eS_{i\pm}\upsilon_{ix}\gamma_{\bot}
\end{equation}
where $S_{\pm} = S_{x} \pm i S_{y}$, $\upsilon_{x}$ is the $x$-component of the velocity and $\gamma_{\bot}$ is in
plane spin-Coulomb drag coefficient. The Fourier transformation of
Eq.~(\ref{eq_motion_transverse1}) leads to the following formula
for the left and right circularly polarized spin current densities:
\begin{equation}\label{j_+/-}
i(\omega \pm \omega_0)j_{\pm} =-ne^2\frac{E}{m^*}
+\frac{j_{\pm}}{\tau} + j_{\pm}\gamma_{\bot}
\end{equation}
where we used the fact that in the absence of SO coupling:
\begin{equation}\label{eq_motion_transverse2}
\frac{dS_{\pm}}{dt}=-i[S_{\pm},\hat{H}]=\pm i\omega_0S_{\pm}
\end{equation}
Then the circularly polarized spin conductivities have the form:
\begin{equation}\label{sigma_circular}
\sigma_{\pm}=\frac{ne^2}{m^*}\frac{1}{-(\omega\pm\omega_0)i+\frac{1}{\tau}+\gamma_{\bot}}
\end{equation}
where we assumed that the spin-Coulomb drag coefficient is the
same for both circular polarizations.\footnote{It is certainly
true in a limit of $\omega\rightarrow 0$}  Accordingly, the
real parts of spin conductivities have the following form:
\begin{equation}\label{sigma_s}
\Re e\sigma^s_{\pm}(\omega)
=\frac{ne^2}{m^*}\frac{\gamma_{\bot}+1/\tau}{(\omega\mp\omega_0)^2
+(\gamma_{\bot} + 1/\tau)^2}
\end{equation}
Assuming that $\omega \gg 1/\tau,\gamma,\omega_0$ we find the
real part of the  transverse spin channel conductivity
$\sigma^s_{\bot}(\omega)$ as follows: Eq.~(\ref{sigma_s}):
\begin{eqnarray}\label{Resigma_s}
\Re e \sigma^s_{\bot}(\omega)= \frac{\Re e
\sigma^s_{+}(\omega)+\Re e \sigma^s_{-}(\omega)}{4} =
\frac{ne^2}{2m^*}\frac{\gamma_{\bot}+1/\tau} {\omega^2}~.
\end{eqnarray}
Eq.~(\ref{Resigma_s}) separates the disorder and interaction
terms. Since we have already included disorder in the
phenomenological Eq.~(\ref{j_+/-}), it is justified to derive the
transverse spin-drag coefficient $\gamma_{\bot}$ by comparing
Eq.~(\ref{Resigma_s}) with the transverse spin-channel
conductivity found from Eq.~(\ref{sigmasperp}).
Using two times the identity~(\ref{eom}) we can rewrite
Eq.~(\ref{sigmasperp}) in terms of the force-force correlation function:
\begin{equation}\label{force-force}
Re\sigma_{\bot}(\omega)= - \frac{4e^2}{m^2 {\cal V}} \Im m
\frac{\langle \langle\sum_n \hat S_{nx}\hat F_{nx};\sum_n \hat
S_{nx}\hat F_{nx}\rangle\rangle_\omega}{\omega^3}
\end{equation}
Then by comparing Eq.~(\ref{Resigma_s}) with
Eq.~(\ref{force-force}) we obtain in the clean limit:
\begin{widetext}
\begin{eqnarray}\label{ReSigmaxx}
&&\gamma_{\bot}(\omega)= - \frac{8}{n m^* {\cal V}} \Im m
\frac{\langle \langle\sum_n \hat S_{nx}\hat F_{nx};\sum_n \hat
S_{nx}\hat F_{nx}\rangle\rangle_\omega}{\omega}= \nonumber\\ &&
\frac{1}{2\pi^2 n m^*} \int_{0}^{\infty}dq q^3\nu^2_q
\frac{e^{-\beta\omega}-1}{\omega}\int_0^\omega d\omega' \frac{\Im
m\chi_{nn}(q,\omega') \Im m
\chi_{xx}(q,\omega-\omega')}{(e^{-\beta\omega'}-1)(e^{-\beta(\omega-\omega')}-1)}=
\nonumber\\ && \frac{k_F^4}{2\pi^2n m^*}\int_0^{\infty}
d\bar{q}\bar{q}^3\nu^2_{\bar{q}}\int_0^{\bar{\omega}}\frac{e^{-2\beta
E_F \bar{\bar{\omega}}}-1}{\bar{\omega}} d\bar{\omega}'\frac{\Im
m\chi_{nn}(\bar{q},\bar{\omega}')\Im
m\chi_{xx}(\bar{q},\bar{\omega}-\bar{\omega}')}{(e^{-2\beta
E_F\bar{\omega}'}-1)(e^{-2\beta
E_F(\bar{\omega}-\bar{\omega}')}-1)}
\end{eqnarray}
\end{widetext}
where $\bar{q}=q/k_F$, $\bar{\omega}=\hbar\omega/2E_F$. For small
$\bar{\omega}$
\begin{equation}
\nu_{\bar{q}}\Im m
\chi_{xx}(\bar{q},\bar{\omega})=-\frac{1}{\sqrt{2}}\frac{\bar\omega/\bar
q}{\sqrt{1-\frac{p^2}{\bar q^2}-\frac{\bar
q^2}{4}}}\frac{r_s}{\bar{q}+r_s\sqrt{2}}
\end{equation}
and $\Im m \chi_{nn}(\bar{q},\bar{\omega}) =\Im m
\chi_{\uparrow}(\bar{q},\bar{\omega})+\Im m \chi_{\downarrow}(\bar{q},\bar{\omega})$, where
\begin{equation}
\nu_{\bar{q}}\Im m \chi_{\uparrow}(\bar{q},\bar{\omega}) =
-\frac{1}{\sqrt{2}}\frac{\bar\omega/\bar q}{\sqrt{1+p-\frac{\bar
q^2}{4}}}\frac{r_s}{\bar{q}+r_s\sqrt{2}}
\end{equation}
and $\Im m \chi_{\downarrow}(\bar{q},\bar{\omega})$ is obtained
from the above equation simply by changing the sign of $p$.
$\nu_{\bar{q}}$ is the dimensionless screened interaction
potential and $r_s=\frac{1}{\sqrt{\pi n}a_B}$ is the dimensionless
Wigner-Seitz radius. In the limit of $T/T_F \ll 1$ this formula
can be evaluated analytically leading to the following integral on
the transverse spin Coulomb drag coefficient:
\begin{widetext}
%\begin{eqnarray}\label{Resigmaxx_final}
%\gamma_{\bot}(0)=\frac{E_F}{\hbar}\left(\frac{k_BT}{E_F}\right)^2\frac{\pi
%r_s^2}{68 }\left[dx \int_{b(p)}^c f(x,p)+\int_{b(-p)}^1 f(x,-p)
%\right]\nonumber\\
%\end{eqnarray}
\begin{equation}
\gamma_{\bot}(0)=\frac{E_F}{\hbar}\left(\frac{k_BT}{E_F}\right)^2\frac{\pi
r_s^2}{6}\left[F(r_s,p)+F(r_s,-p)\right]
\end{equation}
where
%\begin{eqnarray}
%&&f(x,p)= \Im m\chi_{\uparrow}(x)\Im m\chi_{xx}(x)=\nonumber\\ &&
%\frac{1}{\sqrt{1-x}(\sqrt{x}+\frac{r_s}{\sqrt{2(1+p)}})^2}
%\left[\frac{p/[x(1+p)]+2}{\sqrt{(1+p)^2(1-x)-
%\frac{p^2}{4x}-p(1+p)}}
%+\frac{-p/[x(1+p)]+2}{\sqrt{(1+p)[1-x(1+p)]-\frac{p^2}{4x}}}\right]\nonumber\\
%\end{eqnarray}
\begin{equation}\label{Resigmaxx_final}
F(r_s,p) = \int_{a(p)}^{b(p)}
\frac{\bar{q}d\bar{q}}{(\bar{q}+\sqrt{2}r_s)^2}
\frac{1}{\sqrt{\left(1-\frac{p^2}{\bar{q}^2}-\frac{\bar{q}^2}{4}\right)\left(1+p-\frac{\bar{q}^2}{4}\right)}}~,
\end{equation}
for $b(p) \geq a(p)$ and zero otherwise, and the limits of integration are
\begin{equation}
a(p) \equiv \vert\sqrt{1+p}-\sqrt{1-p}\vert~,~~~~~~~~~ b(p) \equiv {\rm min}(\sqrt{1+p}+\sqrt{1-p},2\sqrt{1+p})~.
\end{equation}
\end{widetext}

\begin{figure}[thb]
\vskip 0.27 in
\includegraphics[width=3.5in]{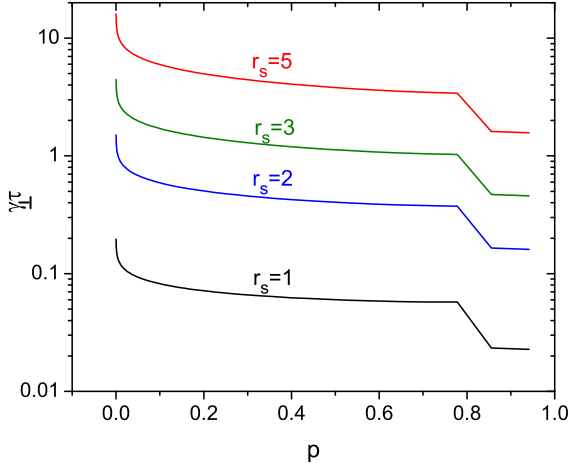}
\caption{The behavior of $\gamma_{\bot}\tau$ as a function of
polarization for various $r_s$. We assume $E_F \tau/\hbar
=10/r_s^2$ and $k_BT/E_F = r_s^2/10$}
\end{figure}

For zero polarization the transverse spin-Coulomb drag coefficient
coincides with the longitudinal one because of the isotropy of the
electron gas. The behavior of the transverse spin-Coulomb drag
coefficient is presented in Fig.~8 for various values of $r_s$.
The effect of the electron-electron interactions is stronger for
larger $r_s$ and therefore $\gamma_{\bot}\tau$ is larger. For a
given $r_s$, $\gamma_{\bot}$ decreases with increasing
polarization and saturates for large polarizations to the value of
$\gamma\tau_{\bot}$ corresponding to the value of the $F(x,p)$
integral for $p=1$ (see Eq.~(\ref{Resigmaxx_final})). The downward
step at polarization $p = 0.8$ is due to the fact that the limits
of integration for $F(r_s,-p)$ come together at this value of $p$
and furthermore the integrand has a singularity that causes the
integral to drop to zero discontinuously at this point.

\section {Appendix D: Direct Matsubara calculation of the
noninteracting transverse spin response}

The Gilbert damping constant  is directly related to
the imaginary part of the transverse spin susceptibility, which does not depend on
the equilibrium parameters (for example magnetization) of the
system. It is useful to show that the calculation of the
transverse spin susceptibility via the torque-torque correlator and the more direct calculation by Matsubara approach are equivalent. We consider the same
Hamiltonian as in Eq.~(\ref{modelHamiltonian}). The transverse
spin response $\chi_{-+}$ is defined as follows:
\begin{equation}
\chi_{-+}(\mathbf{r},\mathbf{r^\prime};t,t^\prime)=\frac{ig^2}{2V}\Theta(t-t^\prime)
\left\langle\left[s_-(\mathbf{r},t),s_+(\mathbf{r^\prime},t^\prime)\right]\right\rangle\,,
\end{equation}
where
$\mathbf{s}(\mathbf{r})=\mbox{Tr}[\boldsymbol{\hat{\sigma}}\hat{\rho}(\mathbf{r})/4]$
is the spin density ($s_\pm=s_x\pm is_y$) in terms of the density
matrix
$\rho_{\alpha\beta}(\mathbf{r})=\Psi^\dagger_\beta(\mathbf{r})\Psi_\alpha(\mathbf{r})$.

It is straightforward to show that the disorder-averaged retarded
single-particle Green's function in the representation of Rashba
subbands  is given by
\begin{equation}
\hat{G}^{R}(\mathbf{k},\mathbf{k^\prime};\omega)=\delta_{\mathbf{k},
\mathbf{k^\prime}}\hat{U}_\mathbf{k}\left(\begin{array}{cc}G_+^R(\mathbf{k},\omega)
& 0\\ 0 &
G_-^R(\mathbf{k},\omega)\end{array}\right)\hat{U}_\mathbf{k}\,,
\end{equation}
where
\begin{equation}
G_\pm^R(\mathbf{k},\omega)=\frac{1}{\omega-(\epsilon_k\pm\\\Omega)+i/2\tau}\,,
\end{equation}
$\epsilon_k=\hbar^2k^2/2m-\mu$\,, and $\hat{U}_\mathbf{k}$ are
spin-rotation matrices:
\begin{equation}
\hat{U}_{\mathbf{k}}=\left(\begin{array}{cc}\cos(\delta/2) &
i\sin(\delta/2)e^{-i\varphi_\mathbf{k}}\\
-i\sin(\delta/2)e^{i\varphi_\mathbf{k}} &
-\cos(\delta/2)\end{array}\right)\,, \label{Uk}
\end{equation}
where
\begin{equation}
\cos \delta \equiv \frac{\omega_0}{\Omega} =
\frac{p}{\sqrt{p^2+\bar\alpha^2}}~,
\end{equation}
and
\begin{equation}
\Omega\equiv \sqrt{\omega_0^2+\alpha^2k_F^2}
\end{equation}
are defined as in main text.  Note that, in general,
$\hat{G}^A=(\hat{G}^R)^\dagger$, in order to obtain the advanced
Green's function.

We do not assume any \textit{a priori} hierarchy in the three
energy scales $H$, $\alpha k_F$, and $1/\tau$, but we
consider for simplicity the limit $\epsilon/\epsilon_F\ll1$, where
$\epsilon$ is any of the mentioned energy scales. Disregarding
terms of order $\epsilon/\epsilon_F$, $k$ can be set to $k_F$ in the
relevant expressions.

It is most convenient to calculate the spin response function in
the Matsubara formalism. For a uniform perturbation, one gets in
general
\begin{align}
\chi_{-+}(i\bar\Omega_n)=-\frac{g^2T}{2V^2}\sum_{\mathbf{k},\mathbf{k^\prime};m}&
G_{\uparrow\uparrow}(\mathbf{k^\prime},\mathbf{k};i\omega_m)\nonumber\\
&\times
G_{\downarrow\downarrow}(\mathbf{k},\mathbf{k^\prime};i\omega_m+i\bar\Omega_n)\,,
\label{chiM}
\end{align}
where $\omega_m=2\pi T(m+1/2)$ and $\bar\Omega_n=2\pi Tn$ are
fermionic and bosonic Matsubara frequencies, respectively. $T$ is
the absolute temperature and $V$ the total volume of the system.
The desired retarded response function is obtained by analytic
continuation $i\bar\Omega_n\to \omega+i0^+$.

It is important to take into account vertex corrections when
averaging Eq.~(\ref{chiM}) over disorder. Otherwise, the response
would have a low-frequency dissipative component even in the
absence of SO interaction, i.e., $\alpha=0$, if the
disorder-averaged Green's functions where inserted in
Eq.~(\ref{chiM}). Namely, we find without \textit{vertex
corrections}:
\begin{equation}
\left.\partial_\omega\mbox{Im}\chi_{-+}(\omega)\right|_{\omega=0}=\frac{g^2}{V}\frac{m^\ast\tau}{8\pi}
\left[\frac{1+\cos^2\delta}{1+(\Omega\tau)^2}+\sin^2\delta\right]\,,
\end{equation}
which diverges in the clean limit, $\tau\to\infty$, even when
there is no SO interaction, i.e., $\delta=0$, and vanishes
in the dirty limit, $\tau\to0$.

Taking into account vertex corrections is conceptually
straightforward but technically somewhat tedious. Defining the
tensor $\tensor{\Pi}$:

\begin{align}
\Pi_{\alpha^\prime\beta,\alpha\beta^\prime}(i\bar\Omega_n)=\sum_{\mathbf{k}\mathbf{k^\prime},m}
&G_{\alpha^\prime\beta}(\mathbf{k^\prime},\mathbf{k};i\omega_m)\nonumber\\
&\times
G_{\alpha\beta^\prime}(\mathbf{k},\mathbf{k^\prime};i\omega_m+i\bar\Omega_n)\,,
\end{align}
whose $\uparrow\uparrow,\downarrow\downarrow$ component determines
the response function according to Eq.~(\ref{chiM}), namely,
$\chi_{-+}=-(T/2V)\Pi_{\uparrow\uparrow,\downarrow\downarrow}$,we
find by summing the ladder diagrams
\begin{align}
\Pi_{\alpha^\prime\beta,\alpha\beta^\prime}(i\bar\Omega_n)=\sum_m&\Gamma_{\alpha^\prime
i,j\beta^\prime}(i\bar\Omega_n,i\omega_m)\nonumber\\
&\times\Upsilon_{i\beta,\alpha j}(i\bar\Omega_n,i\omega_m)\,,
\label{tP}
\end{align}
where
\begin{align}
\Upsilon_{\alpha^\prime\beta,\alpha\beta^\prime}(i\bar\Omega_n,i\omega_m)=
\sum_\mathbf{k}&G_{\alpha^\prime\beta}(\mathbf{k};i\omega_m)\nonumber\\
&\times
G_{\alpha\beta^\prime}(\mathbf{k};i\omega_m+i\bar\Omega_n)\,,
\label{tU}
\end{align}
is the $\omega_m$ contribution to the $\tensor{\Pi}$ tensor
without the vertex corrections and
\begin{align}
\Gamma_{\alpha^\prime\beta,\alpha\beta^\prime}&(i\bar\Omega_n,i\omega_m)=
\delta_{\alpha^\prime\beta}\delta_{\alpha\beta^\prime}+\frac{1}{2\pi\nu\tau}\nonumber\\
&\times\Gamma_{i\beta,\alpha
j}(i\bar\Omega_n,i\omega_m)\Upsilon_{\alpha^\prime
i,j\beta^\prime}(i\bar\Omega_n,i\omega_m) \label{tG}
\end{align}
is the recursive relation for the vertex tensor $\tensor{\Gamma}$.
$\nu=m^\ast/2\pi$ is the density of states per spin. If a tensor
$\tensor{A}$ can be decomposed into matrices $\hat{B}$ and
$\hat{C}$ by
$A_{\alpha^\prime\beta,\alpha\beta^\prime}=B_{\alpha^\prime\beta}C_{\alpha\beta^\prime}$,
I will write $\tensor{A}=\hat{A}\otimes\hat{B}$. All subscripts in
Eqs.~(\ref{tP}), (\ref{tU}), and (\ref{tG}) are spin-$1/2$ indices
taking values $\uparrow$ and $\downarrow$, and the summation over
repeated indices $i$ and $j$ is implied.

In order to calculate the $\tensor{\Pi}$ tensor (\ref{tP}), we
first follow the standard procedure  of replacing the sum over $m$ by
the integral $\int_Cdz\tanh(z/2T)$ with the contour $C$ around the
imaginary axis. The contour is then deformed towards the
$\pm\infty$ along the real axis, around the branch cuts at
$\mbox{Im}z=0$ and $i\bar\Omega_n$. Performing integrals over the
branch cuts, we get
\begin{align}
&\tensor{\Pi}(\omega)=\int_{-\infty}^\infty\frac{d\epsilon}{4\pi
iT\xi}\tanh\left(\frac{\epsilon}{2T}\right)\left[\tensor{\chi}^{R,R}(\epsilon,\epsilon+\omega)\right.\nonumber\\
&\hspace{-0.4cm}\left.-\tensor{\chi}^{A,R}(\epsilon,\epsilon+\omega)+\tensor{\chi}^{A,R}(\epsilon-\omega,\epsilon)-\tensor{\chi}^{A,A}(\epsilon-\omega,\epsilon)\right]\,,
\label{Pi}
\end{align}
where
\begin{align}
\tensor{\chi}^{X,Y}&(\epsilon,\epsilon^\prime)=\tensor{\chi}^{X,Y}_0(\epsilon,\epsilon^\prime)\nonumber\\
&+\xi\sum_\mathbf{k}\hat{G}^X(\mathbf{k},\epsilon)\tensor{\chi}^{X,Y}(\epsilon,\epsilon^\prime)\hat{G}^Y(\mathbf{k},\epsilon^\prime)
\label{chiXY}
\end{align}
is the recursion relation corresponding to Eq.~(\ref{tG})
($X,Y=A~{\rm or}~R$ and $\xi^{-1}=2\pi\nu\tau V$) and
\begin{equation}
\tensor{\chi}^{X,Y}_0(\epsilon,\epsilon^\prime)=\xi\sum_\mathbf{k}\hat{G}^X(\mathbf{k},\epsilon)\otimes\hat{G}^Y(\mathbf{k},\epsilon^\prime)\,.
\label{chi0XY}
\end{equation}
Since we are interested in dissipation, let us define
$2\mbox{Im}\tensor{\Pi}=-i(\tensor{\Pi}-\tensor{\Pi}^\dagger)$,
where
$\{{\tensor{\Pi}^\dagger}\}_{\alpha^\prime\beta,\alpha\beta^\prime}=\Pi_{\beta\alpha^\prime,\beta^\prime\alpha}^\ast$.
We then take the following steps in order to evaluate
$\mbox{Im}\tensor{\Pi}$: First, Eq.~(\ref{chiXY}) is iteratively
expanded in terms of the disorder-averaged Green's functions
$\hat{G}^X$, then the ``unrolled" $\tensor{\chi}^{X,Y}$ is
substituted into Eq.~(\ref{Pi}). Differentiating the resulting
expression with respect to $\omega$ and taking the imaginary part,
the integral can be transformed integrating by parts into $\int
d\epsilon\tanh(\epsilon/2T)\partial_\epsilon...=-\int
d\epsilon\partial_\epsilon\tanh(\epsilon/2T)...$, and, assuming
low temperatures, we approximate
$\partial_\epsilon\tanh(\epsilon/2T)\approx2\delta(\epsilon)$, so
that the dissipation is naturally governed by
electron-hole pair excitations near the Fermi surface. The infinite summation
series is finally ``rolled" back into a recursive relation and we
obtain the following expression for the spin response:
\begin{align}
\left.\partial_\omega\mbox{Im}
\chi_{-+}(\omega)\right|_{\omega=0}=&-\frac{g^2}{V}\frac{m^\ast\tau}{8\pi}\nonumber\\
&\hspace{-2cm}\times\left(\tensor{\chi}^{R,R}-\tensor{\chi}^{A,R}-\tensor{\chi}^{R,A}
+\tensor{\chi}^{A,A}\right)_{\uparrow\uparrow,\downarrow\downarrow}\,,
\end{align}
where all $\tensor{\chi}$'s are now evaluated at
$\epsilon,\epsilon^\prime=0$.

The problem is thus reduced to calculating
$\tensor{\chi}^{X,Y}(0,0)$ using Eqs.~(\ref{chiXY}) and
(\ref{chi0XY}). It is still somewhat tedious but now totally
straightforward. We first do the angular integration fixing the
absolute value of $\mathbf{k}$ in Eq.~(\ref{chi0XY}). This
averages over the rotation matrices (\ref{Uk}). The resulting
angle-averaged tensor (\ref{chi0XY}) then decomposes into 6
components: $\hat{1}\otimes\hat{1}$,
$\hat{1}\otimes\hat{\sigma}_z$, $\hat{\sigma}_z\otimes\hat{1}$,
$\hat{\sigma}_z\otimes\hat{\sigma}_z$,
$\hat{\sigma}_+\otimes\hat{\sigma}_-$, and
$\hat{\sigma}_-\otimes\hat{\sigma}_+$ (the last two are solely due
to the SO coupling), where
$\hat{\sigma}_\pm=\hat{\sigma}_x\pm i\hat{\sigma}_y$. The
respective prefactors are given by integrals over the absolute
value of momentum near the Fermi energy, which are trivial to
evaluate after linearizing the dispersion at the Fermi level. We
then make an ansatz that the tensor $\tensor{\chi}^{X,Y}(0,0)$
determined by Eq.~(\ref{chiXY}) can also be expanded in terms of
the same 6 components and, after plugging the expanded
$\tensor{\chi}^{X,Y}$ (with unknown coefficients) and
$\tensor{\chi}^{X,Y}_0$ (with calculated coefficients) into
Eq.~(\ref{chiXY}), we obtain a linear system of equations for the
unknown coefficients that determine $\tensor{\chi}^{X,Y}$. Solving
this (with Mathematica), we find a solution (validating the
ansatz), and the dissipative part of the spin response is finally
given by
\begin{align}
\left.\partial_\omega\mbox{Im}\chi_{-+}(\omega)\right|_{\omega=0}=
&\frac{g^2}{V}\frac{m^\ast\tau}{8\pi}\sin^2\delta\nonumber\\
&\times\frac{1+\frac{1}{2}(\Omega\tau)^2(1+\cos^2\delta)}{\cos^2\delta+\frac{1}{4}(\Omega\tau)^2(1+\cos^2\delta)^2}\,,
\label{chi}
\end{align}
Using Eq.~(\ref{chi_matrix}) and Eq.~(\ref{GilbertMatrix}) one can
show that Gilbert damping has a following form:
\begin{eqnarray}
\lambda =gM_0V \lim_{\omega \rightarrow 0}
\frac{\partial_\omega\mbox{Im}\chi_{-+}(\omega)}{\chi^2_0(\omega=0)}=\nonumber\\
p\frac{ E_F\tau\sin^2\delta}{\hbar}\frac{1+\frac{1}{2}
(\Omega\tau)^2(1+\cos^2\delta)}{\cos^2\delta+\frac{1}{4}(\Omega\tau)^2(1+\cos^2\delta)^2}~
\end{eqnarray}
in agreement with Eq.~(\ref{lambdani}).
%\bibliography{Gilbertoct03}

\begin{thebibliography}{34}
\expandafter\ifx\csname
natexlab\endcsname\relax\def\natexlab#1{#1}\fi
\expandafter\ifx\csname bibnamefont\endcsname\relax
  \def\bibnamefont#1{#1}\fi
\expandafter\ifx\csname bibfnamefont\endcsname\relax
  \def\bibfnamefont#1{#1}\fi
\expandafter\ifx\csname citenamefont\endcsname\relax
  \def\citenamefont#1{#1}\fi
\expandafter\ifx\csname url\endcsname\relax
  \def\url#1{\texttt{#1}}\fi
\expandafter\ifx\csname
urlprefix\endcsname\relax\def\urlprefix{URL }\fi
\providecommand{\bibinfo}[2]{#2}
\providecommand{\eprint}[2][]{\url{#2}}

\bibitem[{\citenamefont{Gilbert}(1955)}]{Gilbert55}
\bibinfo{author}{\bibfnamefont{T.~L.} \bibnamefont{Gilbert}},
  \bibinfo{journal}{Phys. Rev.} \textbf{\bibinfo{volume}{100}},
  \bibinfo{pages}{1243} (\bibinfo{year}{1955}).

\bibitem[{\citenamefont{Gilbert}(2004)}]{Gilbert04}
\bibinfo{author}{\bibfnamefont{T.~L.} \bibnamefont{Gilbert}},
  \bibinfo{journal}{IEEE Trans. Magn.} \textbf{\bibinfo{volume}{40}},
  \bibinfo{pages}{3443} (\bibinfo{year}{2004}).

\bibitem[{\citenamefont{Lifshitz and Pitaevskii}(1980)}]{landauBOOKv9}
\bibinfo{author}{\bibfnamefont{E.~M.} \bibnamefont{Lifshitz}} \bibnamefont{and}
  \bibinfo{author}{\bibfnamefont{L.~P.} \bibnamefont{Pitaevskii}},
  \emph{\bibinfo{title}{Statistical Physics, Part 2}}, vol.~\bibinfo{volume}{9}
  of \emph{\bibinfo{series}{Course of Theoretical Physics}}
  (\bibinfo{publisher}{Pergamon}, \bibinfo{address}{Oxford},
  \bibinfo{year}{1980}), \bibinfo{edition}{3rd} ed.

\bibitem[{\citenamefont{Tserkovnyak et~al.}(2005)\citenamefont{Tserkovnyak,
  Brataas, Bauer, and Halperin}}]{Yaroslav05}
\bibinfo{author}{\bibfnamefont{Y.}~\bibnamefont{Tserkovnyak}},
  \bibinfo{author}{\bibfnamefont{A.}~\bibnamefont{Brataas}},
  \bibinfo{author}{\bibfnamefont{G.~E.} \bibnamefont{Bauer}}, \bibnamefont{and}
  \bibinfo{author}{\bibfnamefont{B.~I.} \bibnamefont{Halperin}},
  \bibinfo{journal}{Rev. Mod. Phys.} \textbf{\bibinfo{volume}{77}},
  \bibinfo{pages}{1375} (\bibinfo{year}{2005}).

\bibitem[{\citenamefont{Tserkovnyak
  et~al.}(2002{\natexlab{a}})\citenamefont{Tserkovnyak, Brataas, and
  Bauer}}]{tserkovPRL02sp}
\bibinfo{author}{\bibfnamefont{Y.}~\bibnamefont{Tserkovnyak}},
  \bibinfo{author}{\bibfnamefont{A.}~\bibnamefont{Brataas}}, \bibnamefont{and}
  \bibinfo{author}{\bibfnamefont{G.~E.~W.} \bibnamefont{Bauer}},
  \bibinfo{journal}{Phys. Rev. Lett.} \textbf{\bibinfo{volume}{88}},
  \bibinfo{eid}{117601} (\bibinfo{year}{2002}{\natexlab{a}}).

\bibitem[{\citenamefont{Tserkovnyak
  et~al.}(2002{\natexlab{b}})\citenamefont{Tserkovnyak, Brataas, and
  Bauer}}]{tserkovPRB02sp}
\bibinfo{author}{\bibfnamefont{Y.}~\bibnamefont{Tserkovnyak}},
  \bibinfo{author}{\bibfnamefont{A.}~\bibnamefont{Brataas}}, \bibnamefont{and}
  \bibinfo{author}{\bibfnamefont{G.~E.~W.} \bibnamefont{Bauer}},
  \bibinfo{journal}{Phys. Rev. B} \textbf{\bibinfo{volume}{66}},
  \bibinfo{eid}{224403} (\bibinfo{year}{2002}{\natexlab{b}}).

\bibitem[{\citenamefont{Heinrich et~al.}(1967)\citenamefont{Heinrich,
  Fraitov{\'{a}}, and Kambersk{\'{y}}}}]{heinrichPSS67}
\bibinfo{author}{\bibfnamefont{B.}~\bibnamefont{Heinrich}},
  \bibinfo{author}{\bibfnamefont{D.}~\bibnamefont{Fraitov{\'{a}}}},
  \bibnamefont{and}
  \bibinfo{author}{\bibfnamefont{V.}~\bibnamefont{Kambersk{\'{y}}}},
  \bibinfo{journal}{Phys. Status Solidi} \textbf{\bibinfo{volume}{23}},
  \bibinfo{pages}{501} (\bibinfo{year}{1967}).

\bibitem[{\citenamefont{Korenman and Prange}(1972)}]{Korenman72}
\bibinfo{author}{\bibfnamefont{V.}~\bibnamefont{Korenman}} \bibnamefont{and}
  \bibinfo{author}{\bibfnamefont{R.~E.} \bibnamefont{Prange}},
  \bibinfo{journal}{Phys. Rev. B} \textbf{\bibinfo{volume}{6}},
  \bibinfo{pages}{2769} (\bibinfo{year}{1972}).

\bibitem[{\citenamefont{Lutovinov and Reizer}(1979)}]{lutovinovJETP79}
\bibinfo{author}{\bibfnamefont{V.~S.} \bibnamefont{Lutovinov}}
  \bibnamefont{and} \bibinfo{author}{\bibfnamefont{M.~U.}
  \bibnamefont{Reizer}}, \bibinfo{journal}{Sov. Phys. JETP}
  \textbf{\bibinfo{volume}{50}}, \bibinfo{pages}{355} (\bibinfo{year}{1979}).

\bibitem[{\citenamefont{Solontsov and Vasil'ev}(1993)}]{solontsovPLA93}
\bibinfo{author}{\bibfnamefont{A.~Z.} \bibnamefont{Solontsov}}
  \bibnamefont{and} \bibinfo{author}{\bibfnamefont{A.~N.}
  \bibnamefont{Vasil'ev}}, \bibinfo{journal}{Phys. Lett. A}
  \textbf{\bibinfo{volume}{177}}, \bibinfo{pages}{362} (\bibinfo{year}{1993}).

\bibitem[{\citenamefont{Suhl}(1998)}]{suhlIEEEM98}
\bibinfo{author}{\bibfnamefont{H.}~\bibnamefont{Suhl}}, \bibinfo{journal}{IEEE
  Trans. Magn.} \textbf{\bibinfo{volume}{34}}, \bibinfo{pages}{1834}
  (\bibinfo{year}{1998}).

\bibitem[{\citenamefont{Heinrich et~al.}(2002)\citenamefont{Heinrich, Urban,
  and Woltersdorf}}]{heinrichIEEEM02}
\bibinfo{author}{\bibfnamefont{B.}~\bibnamefont{Heinrich}},
  \bibinfo{author}{\bibfnamefont{R.}~\bibnamefont{Urban}}, \bibnamefont{and}
  \bibinfo{author}{\bibfnamefont{G.}~\bibnamefont{Woltersdorf}},
  \bibinfo{journal}{IEEE Trans. Magn.} \textbf{\bibinfo{volume}{38}},
  \bibinfo{pages}{2496} (\bibinfo{year}{2002}).

\bibitem[{\citenamefont{Kune{\v{s}} and Kambersk{\'{y}}}(2002)}]{kunesPRB02}
\bibinfo{author}{\bibfnamefont{J.}~\bibnamefont{Kune{\v{s}}}} \bibnamefont{and}
  \bibinfo{author}{\bibfnamefont{V.}~\bibnamefont{Kambersk{\'{y}}}},
  \bibinfo{journal}{Phys. Rev. B} \textbf{\bibinfo{volume}{65}},
  \bibinfo{eid}{212411} (\bibinfo{year}{2002}).

\bibitem[{\citenamefont{Dobin and Victora}(2003)}]{dobinPRL03}
\bibinfo{author}{\bibfnamefont{A.~Y.} \bibnamefont{Dobin}} \bibnamefont{and}
  \bibinfo{author}{\bibfnamefont{R.~H.} \bibnamefont{Victora}},
  \bibinfo{journal}{Phys. Rev. Lett.} \textbf{\bibinfo{volume}{90}},
  \bibinfo{eid}{167203} (\bibinfo{year}{2003}).

\bibitem[{\citenamefont{Sinova et~al.}(2004{\natexlab{a}})\citenamefont{Sinova,
  Jungwirth, Liu, Sasaki, Furdyna, Atkinson, and MacDonald}}]{sinovaPRB04}
\bibinfo{author}{\bibfnamefont{J.}~\bibnamefont{Sinova}},
  \bibinfo{author}{\bibfnamefont{T.}~\bibnamefont{Jungwirth}},
  \bibinfo{author}{\bibfnamefont{X.}~\bibnamefont{Liu}},
  \bibinfo{author}{\bibfnamefont{Y.}~\bibnamefont{Sasaki}},
  \bibinfo{author}{\bibfnamefont{J.~K.} \bibnamefont{Furdyna}},
  \bibinfo{author}{\bibfnamefont{W.~A.} \bibnamefont{Atkinson}},
  \bibnamefont{and} \bibinfo{author}{\bibfnamefont{A.~H.}
  \bibnamefont{MacDonald}}, \bibinfo{journal}{Phys. Rev. B}
  \textbf{\bibinfo{volume}{69}}, \bibinfo{eid}{085209}
  (\bibinfo{year}{2004}{\natexlab{a}}).

\bibitem[{\citenamefont{Tserkovnyak et~al.}(2004)\citenamefont{Tserkovnyak,
  Fiete, and Halperin}}]{Tserkovnyak04}
\bibinfo{author}{\bibfnamefont{Y.}~\bibnamefont{Tserkovnyak}},
  \bibinfo{author}{\bibfnamefont{G.~A.} \bibnamefont{Fiete}}, \bibnamefont{and}
  \bibinfo{author}{\bibfnamefont{B.~I.} \bibnamefont{Halperin}},
  \bibinfo{journal}{Apl. Phys. Lett.} \textbf{\bibinfo{volume}{84}},
  \bibinfo{pages}{5234} (\bibinfo{year}{2004}).

\bibitem[{\citenamefont{Koopmans et~al.}(2005)\citenamefont{Koopmans, Ruigrok,
  {F. Dalla Longa}, and de~Jonge}}]{koopmansPRL05}
\bibinfo{author}{\bibfnamefont{B.}~\bibnamefont{Koopmans}},
  \bibinfo{author}{\bibfnamefont{J.~J.~M.} \bibnamefont{Ruigrok}},
  \bibinfo{author}{\bibnamefont{{F. Dalla Longa}}}, \bibnamefont{and}
  \bibinfo{author}{\bibfnamefont{W.~J.~M.} \bibnamefont{de~Jonge}},
  \bibinfo{journal}{Phys. Rev. Lett.} \textbf{\bibinfo{volume}{95}},
  \bibinfo{eid}{267207} (\bibinfo{year}{2005}).

\bibitem[{\citenamefont{Steiauf and F{\"{a}}hnle}(2005)}]{steiaufPRB05}
\bibinfo{author}{\bibfnamefont{D.}~\bibnamefont{Steiauf}} \bibnamefont{and}
  \bibinfo{author}{\bibfnamefont{M.}~\bibnamefont{F{\"{a}}hnle}},
  \bibinfo{journal}{Phys. Rev. B} \textbf{\bibinfo{volume}{72}},
  \bibinfo{eid}{064450} (\bibinfo{year}{2005}).

\bibitem[{\citenamefont{Tserkovnyak et~al.}()\citenamefont{Tserkovnyak,
  Skadsem, Brataas, and Bauer}}]{tserkovPRB06}
\bibinfo{author}{\bibfnamefont{Y.}~\bibnamefont{Tserkovnyak}},
  \bibinfo{author}{\bibfnamefont{H.~J.} \bibnamefont{Skadsem}},
  \bibinfo{author}{\bibfnamefont{A.}~\bibnamefont{Brataas}}, \bibnamefont{and}
  \bibinfo{author}{\bibfnamefont{G.~E.~W.} \bibnamefont{Bauer}},
  \bibinfo{note}{cond-mat/0512715}.

\bibitem[{\citenamefont{Kohno et~al.}()\citenamefont{Kohno, Tatara, and
  Shibata}}]{kohnoCM06}
\bibinfo{author}{\bibfnamefont{H.}~\bibnamefont{Kohno}},
  \bibinfo{author}{\bibfnamefont{G.}~\bibnamefont{Tatara}}, \bibnamefont{and}
  \bibinfo{author}{\bibfnamefont{J.}~\bibnamefont{Shibata}},
  \bibinfo{note}{cond-mat/0605186}.

\bibitem[{\citenamefont{Mizukami et~al.}(2002)\citenamefont{Mizukami, Ando, and
  Miyazaki}}]{mizukamiPRB02}
\bibinfo{author}{\bibfnamefont{S.}~\bibnamefont{Mizukami}},
  \bibinfo{author}{\bibfnamefont{Y.}~\bibnamefont{Ando}}, \bibnamefont{and}
  \bibinfo{author}{\bibfnamefont{T.}~\bibnamefont{Miyazaki}},
  \bibinfo{journal}{Phys. Rev. B} \textbf{\bibinfo{volume}{66}},
  \bibinfo{eid}{104413} (\bibinfo{year}{2002}).

\bibitem[{\citenamefont{Ingvarsson et~al.}(2002)\citenamefont{Ingvarsson,
  Ritchie, Liu, Xiao, Slonczewski, Trouilloud, and Koch}}]{ingvarssonPRB02}
\bibinfo{author}{\bibfnamefont{S.}~\bibnamefont{Ingvarsson}},
  \bibinfo{author}{\bibfnamefont{L.}~\bibnamefont{Ritchie}},
  \bibinfo{author}{\bibfnamefont{X.~Y.} \bibnamefont{Liu}},
  \bibinfo{author}{\bibfnamefont{G.}~\bibnamefont{Xiao}},
  \bibinfo{author}{\bibfnamefont{J.~C.} \bibnamefont{Slonczewski}},
  \bibinfo{author}{\bibfnamefont{P.~L.} \bibnamefont{Trouilloud}},
  \bibnamefont{and} \bibinfo{author}{\bibfnamefont{R.~H.} \bibnamefont{Koch}},
  \bibinfo{journal}{Phys. Rev. B} \textbf{\bibinfo{volume}{66}},
  \bibinfo{eid}{214416} (\bibinfo{year}{2002}).

\bibitem[{\citenamefont{Burkov and Balents}(2004)}]{Balents04}
\bibinfo{author}{\bibfnamefont{A.~A.} \bibnamefont{Burkov}} \bibnamefont{and}
  \bibinfo{author}{\bibfnamefont{L.}~\bibnamefont{Balents}},
  \bibinfo{journal}{Phys. Rev. B} \textbf{\bibinfo{volume}{69}},
  \bibinfo{pages}{245312} (\bibinfo{year}{2004}).

\bibitem[{\citenamefont{Murakami et~al.}(2003)\citenamefont{Murakami, Nagaosa,
  and Zhang}}]{Murakami03}
\bibinfo{author}{\bibfnamefont{S.}~\bibnamefont{Murakami}},
  \bibinfo{author}{\bibfnamefont{N.}~\bibnamefont{Nagaosa}}, \bibnamefont{and}
  \bibinfo{author}{\bibfnamefont{S.-C.} \bibnamefont{Zhang}},
  \bibinfo{journal}{Science} \textbf{\bibinfo{volume}{301}},
  \bibinfo{pages}{1348} (\bibinfo{year}{2003}).

\bibitem[{\citenamefont{Sinova et~al.}(2004{\natexlab{b}})\citenamefont{Sinova,
  Culcer, Niu, Sinitsyn, Jungwirth, and MacDonald}}]{Sinova04}
\bibinfo{author}{\bibfnamefont{J.}~\bibnamefont{Sinova}},
  \bibinfo{author}{\bibfnamefont{D.}~\bibnamefont{Culcer}},
  \bibinfo{author}{\bibfnamefont{Q.}~\bibnamefont{Niu}},
  \bibinfo{author}{\bibfnamefont{N.~A.} \bibnamefont{Sinitsyn}},
  \bibinfo{author}{\bibfnamefont{T.}~\bibnamefont{Jungwirth}},
  \bibnamefont{and} \bibinfo{author}{\bibfnamefont{A.~H.}
  \bibnamefont{MacDonald}}, \bibinfo{journal}{Phys. Rev. Lett.}
  \textbf{\bibinfo{volume}{92}}, \bibinfo{pages}{126603}
  (\bibinfo{year}{2004}{\natexlab{b}}).

\bibitem[{\citenamefont{Kato et~al.}(2004)\citenamefont{Kato, Myers, Gossard,
  and Awschalom}}]{Kato04}
\bibinfo{author}{\bibfnamefont{Y.~K.} \bibnamefont{Kato}},
  \bibinfo{author}{\bibfnamefont{R.~C.} \bibnamefont{Myers}},
  \bibinfo{author}{\bibfnamefont{A.~C.} \bibnamefont{Gossard}},
  \bibnamefont{and} \bibinfo{author}{\bibfnamefont{D.~D.}
  \bibnamefont{Awschalom}}, \bibinfo{journal}{Science}
  \textbf{\bibinfo{volume}{306}}, \bibinfo{pages}{1910} (\bibinfo{year}{2004}).

\bibitem[{\citenamefont{Wunderlich et~al.}(2005)\citenamefont{Wunderlich,
  Kaestner, Sinova, and Jungwirth}}]{Wunderlich05}
\bibinfo{author}{\bibfnamefont{J.}~\bibnamefont{Wunderlich}},
  \bibinfo{author}{\bibfnamefont{B.}~\bibnamefont{Kaestner}},
  \bibinfo{author}{\bibfnamefont{J.}~\bibnamefont{Sinova}}, \bibnamefont{and}
  \bibinfo{author}{\bibfnamefont{T.}~\bibnamefont{Jungwirth}},
  \bibinfo{journal}{Phys. Rev. Lett.} \textbf{\bibinfo{volume}{94}},
  \bibinfo{pages}{047204} (\bibinfo{year}{2005}).

\bibitem[{\citenamefont{Sih et~al.}(2006)\citenamefont{Sih, Lau, Myers,
  Horowitz, Gossard, and Awschalom}}]{sihPRL06}
\bibinfo{author}{\bibfnamefont{V.}~\bibnamefont{Sih}},
  \bibinfo{author}{\bibfnamefont{W.~H.} \bibnamefont{Lau}},
  \bibinfo{author}{\bibfnamefont{R.~C.} \bibnamefont{Myers}},
  \bibinfo{author}{\bibfnamefont{V.~R.} \bibnamefont{Horowitz}},
  \bibinfo{author}{\bibfnamefont{A.~C.} \bibnamefont{Gossard}},
  \bibnamefont{and} \bibinfo{author}{\bibfnamefont{D.~D.}
  \bibnamefont{Awschalom}}, \bibinfo{journal}{Phys. Rev. Lett.}
  \textbf{\bibinfo{volume}{97}}, \bibinfo{pages}{096605}
  (\bibinfo{year}{2006}).

\bibitem[{\citenamefont{Konig et~al.}(2006)\citenamefont{Konig, Tschetschetkin,
  Hankiewicz, Sinova, Hock, Daumer, Schafer, Becker, Buhmann, and
  Molenkamp}}]{Konig06}
\bibinfo{author}{\bibfnamefont{M.}~\bibnamefont{Konig}},
  \bibinfo{author}{\bibfnamefont{A.}~\bibnamefont{Tschetschetkin}},
  \bibinfo{author}{\bibfnamefont{E.~M.} \bibnamefont{Hankiewicz}},
  \bibinfo{author}{\bibfnamefont{J.}~\bibnamefont{Sinova}},
  \bibinfo{author}{\bibfnamefont{V.}~\bibnamefont{Hock}},
  \bibinfo{author}{\bibfnamefont{V.}~\bibnamefont{Daumer}},
  \bibinfo{author}{\bibfnamefont{M.}~\bibnamefont{Schafer}},
  \bibinfo{author}{\bibfnamefont{C.~R.} \bibnamefont{Becker}},
  \bibinfo{author}{\bibfnamefont{H.}~\bibnamefont{Buhmann}}, \bibnamefont{and}
  \bibinfo{author}{\bibfnamefont{L.~W.} \bibnamefont{Molenkamp}},
  \bibinfo{journal}{Phys. Rev. Lett.} \textbf{\bibinfo{volume}{96}},
  \bibinfo{pages}{076804} (\bibinfo{year}{2006}).

\bibitem[{\citenamefont{D'Amico and Vignale}(2000)}]{Amico00}
\bibinfo{author}{\bibfnamefont{I.}~\bibnamefont{D'Amico}} \bibnamefont{and}
  \bibinfo{author}{\bibfnamefont{G.}~\bibnamefont{Vignale}},
  \bibinfo{journal}{Phys. Rev. B} \textbf{\bibinfo{volume}{62}},
  \bibinfo{pages}{4853} (\bibinfo{year}{2000}).

\bibitem[{\citenamefont{D'Amico and Vignale}(2002)}]{Amico02}
\bibinfo{author}{\bibfnamefont{I.}~\bibnamefont{D'Amico}} \bibnamefont{and}
  \bibinfo{author}{\bibfnamefont{G.}~\bibnamefont{Vignale}},
  \bibinfo{journal}{Phys. Rev. B} \textbf{\bibinfo{volume}{65}},
  \bibinfo{pages}{85109} (\bibinfo{year}{2002}).

\bibitem[{\citenamefont{Qian and Vignale}(2002)}]{qianPRL02}
\bibinfo{author}{\bibfnamefont{Z.}~\bibnamefont{Qian}} \bibnamefont{and}
  \bibinfo{author}{\bibfnamefont{G.}~\bibnamefont{Vignale}},
  \bibinfo{journal}{Phys. Rev. Lett.} \textbf{\bibinfo{volume}{88}},
  \bibinfo{eid}{056404} (\bibinfo{year}{2002}).

\bibitem[{\citenamefont{Mishchenko et~al.}(2004)\citenamefont{Mishchenko,
  Shytov, and Halperin}}]{Halperin04}
\bibinfo{author}{\bibfnamefont{E.~G.} \bibnamefont{Mishchenko}},
  \bibinfo{author}{\bibfnamefont{A.~V.} \bibnamefont{Shytov}},
  \bibnamefont{and} \bibinfo{author}{\bibfnamefont{B.~I.}
  \bibnamefont{Halperin}}, \bibinfo{journal}{Phys. Rev. Lett.}
  \textbf{\bibinfo{volume}{93}}, \bibinfo{pages}{226602}
  (\bibinfo{year}{2004}).

\bibitem[{\citenamefont{S.~Ingvarsson and Koch}()}]{Ingvarsson04}
\bibinfo{author}{\bibfnamefont{S.~P.~P.} \bibnamefont{S.~Ingvarsson},
  \bibfnamefont{G.~Xiao}} \bibnamefont{and}
  \bibinfo{author}{\bibfnamefont{R.~H.} \bibnamefont{Koch}},
  \eprint{cond-mat/0408608}.

\end{thebibliography}

\end{document}